\documentclass[prl,aps,twocolumn,superscriptaddress,floatfix,nofootinbib]{revtex4-1}
\usepackage{graphicx}
\usepackage{amsmath}
\usepackage{amssymb}
\usepackage{subfigure}
\usepackage{times}

\def\togli#1{}
\def\<{\langle}
\def\>{\rangle}

\usepackage{color}

\usepackage{soul}

\begin{document}

\title{Phase estimation via quantum interferometry for noisy detectors}

\author{Nicol\`{o} Spagnolo} \affiliation{Dipartimento di Fisica,
  Sapienza Universit\`{a} di Roma, piazzale Aldo Moro 5, I-00185 Roma,
  Italy} \affiliation{Consorzio Nazionale Interuniversitario per le
  Scienze Fisiche della Materia, piazzale Aldo Moro 5, I-00185 Roma,
  Italy} \author{Chiara Vitelli} \affiliation{Dipartimento di Fisica,
  Sapienza Universit\`{a} di Roma, piazzale Aldo Moro 5, I-00185 Roma,
  Italy} \author{Vito Giovanni Lucivero} \affiliation{Dipartimento di
  Fisica, Sapienza Universit\`{a} di Roma, piazzale Aldo Moro 5,
  I-00185 Roma, Italy} 
\author{Vittorio Giovannetti} \affiliation{NEST, Scuola Normale
  Superiore and Istituto Nanoscienze-CNR, Piazza dei Cavalieri 7,
  I-56126 Pisa, Italy} \author{Lorenzo Maccone}
\affiliation{Dip.~Fisica ``A.~Volta'', INFN Sez.~Pavia, Universit\`a
  di Pavia, via Bassi 6, I-27100 Pavia, Italy}
\author{Fabio Sciarrino}
\email{fabio.sciarrino@uniroma1.it} 
  \homepage{http://quantumoptics.phys.uniroma1.it}
  \affiliation{Dipartimento di
  Fisica, Sapienza Universit\`{a} di Roma, piazzale Aldo Moro 5,
  I-00185 Roma, Italy} 

\begin{abstract}
  The sensitivity in optical interferometry is strongly affected by
  losses during the signal propagation or at the detection stage. The
  optimal quantum states of the probing signals in the presence of
  loss were recently found. However, in
  many cases of practical interest, their associated accuracy is worse
  than the one obtainable without employing quantum resources
  (e.g.~entanglement and squeezing) but neglecting the detector's loss. 
  Here we detail an experiment that can reach the latter
  even in the presence of imperfect detectors: it employs a
  phase-sensitive amplification of the signals \emph{after} the phase
  sensing, \emph{before} the detection. We experimentally 
  demonstrated the feasibility of a phase estimation experiment able 
  to reach its optimal working regime. Since our method uses
  coherent states as input signals, it is a practical technique that
  can be used for high-sensitivity interferometry and, in contrast to
  the optimal strategies, does not require one to have an exact
  characterization of the loss beforehand.\end{abstract}

\maketitle

From the investigation of fragile biological samples, such as tissues 
\cite{Nasr09} or blood proteins in aqueous buffer solution \cite{Cres11}, to
gravitational wave measurements \cite{Goda08,Abad11}, the estimation of an optical phase $\phi$ 
through interferometric experiments is an ubiquitous technique. For each 
input state of the probe, the maximum accuracy of the process, optimized
over all possible measurement strategies, is provided by the quantum
Fisher information $I^{q}_{\phi}$ through the Quantum Cram{\'e}r-Rao (QCR)
bound~\cite{Hels76,Pari09}. The QCR sets an asymptotically achievable lower bound on
the mean square error of the estimation $\delta \phi \geq (M
I^{q}_{\phi})^{-1/2}$, where $M$ is the number of repeated experiments.  In
the absence of noise and when no quantum effects (like entanglement or
squeezing) are exploited in the probe preparation, the QCR bound
scales as the inverse of the mean photon number, the Standard Quantum
Limit (SQL). Better performances are known to be achievable when using
entangled input signals
\cite{Giov06-Dowl08-mmst-Lee09-Cabl10,Dorn09,Aspa09,Kacp10,datta,Giov11}. However, 
all experiments up to now have been performed using post-selection and cannot claim a
sub-SQL sensitivity \cite{post-sel-prl}. An alternative approach, exploited in
gravitational wave interferometry, relies on combining an intense coherent
beam with squeezed light on a beam-splitter, obtaining an enhancement in
the sensitivity of a constant factor proportional to the squeezing factor
\cite{Gran87,Goda08,Abad11}. Additionally, in the presence of
loss, the SQL can be asymptotically beaten only by a constant factor
\cite{Kolo10,Knys11-davido,Aspa09}, so that sophisticated sub-SQL strategies
\cite{Dang01-Mitc04-Hofm07-Walt04,datta} (implemented up to now only
for few photons) may not be worth the effort. This implies that, for
practical high-sensitivity interferometry, the best resource
exploitation (or, equivalently, the minimally invasive scenarios)
currently entail strategies based on the use of a coherent state
$|\alpha\rangle$, i.e.~a classical signal. Its QCR bound takes the
form $\delta \phi \geqslant (2M \eta \xi
|\alpha|^2)^{-1/2}$, where we consider separately the loss $\mathcal{L}_{\xi}=1-\xi$ in the
sensing stage and the loss $\mathcal{L}_{\eta}=1-\eta$ in the 
overall detection process. Here we present the experimental realization of a robust
phase estimation protocol that improves the above accuracy up to
$\sim(2M \xi |\alpha|^2)^{-1}$, while still using coherent signals as
input. It achieves the SQL of a system only affected by the
propagation loss $\mathcal{L}_{\xi}$, and not by 
the detection stage $\mathcal{L}_{\eta}$.

%
%
\begin{figure*}[ht!]
\centering
\begin{minipage}{0.50\textwidth}
\includegraphics[width=0.99\textwidth]{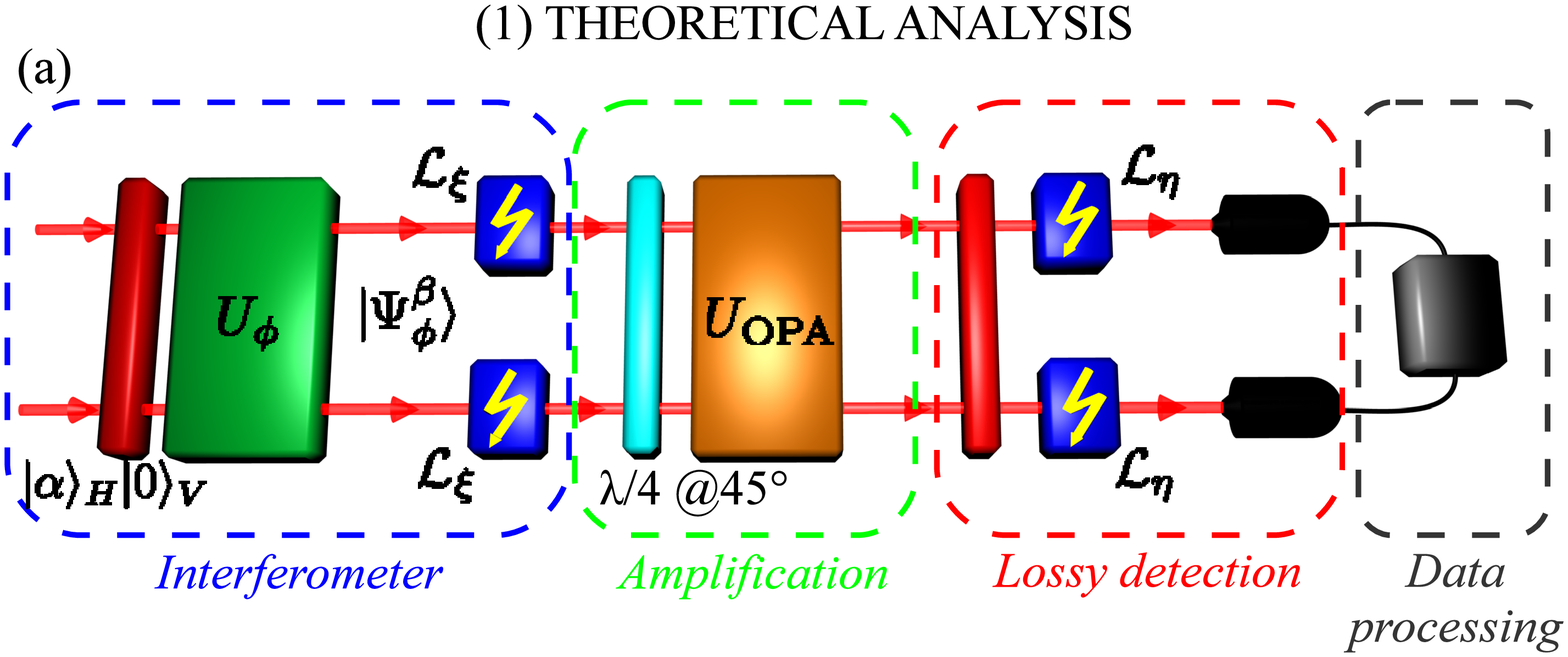}
\includegraphics[width=0.95\textwidth]{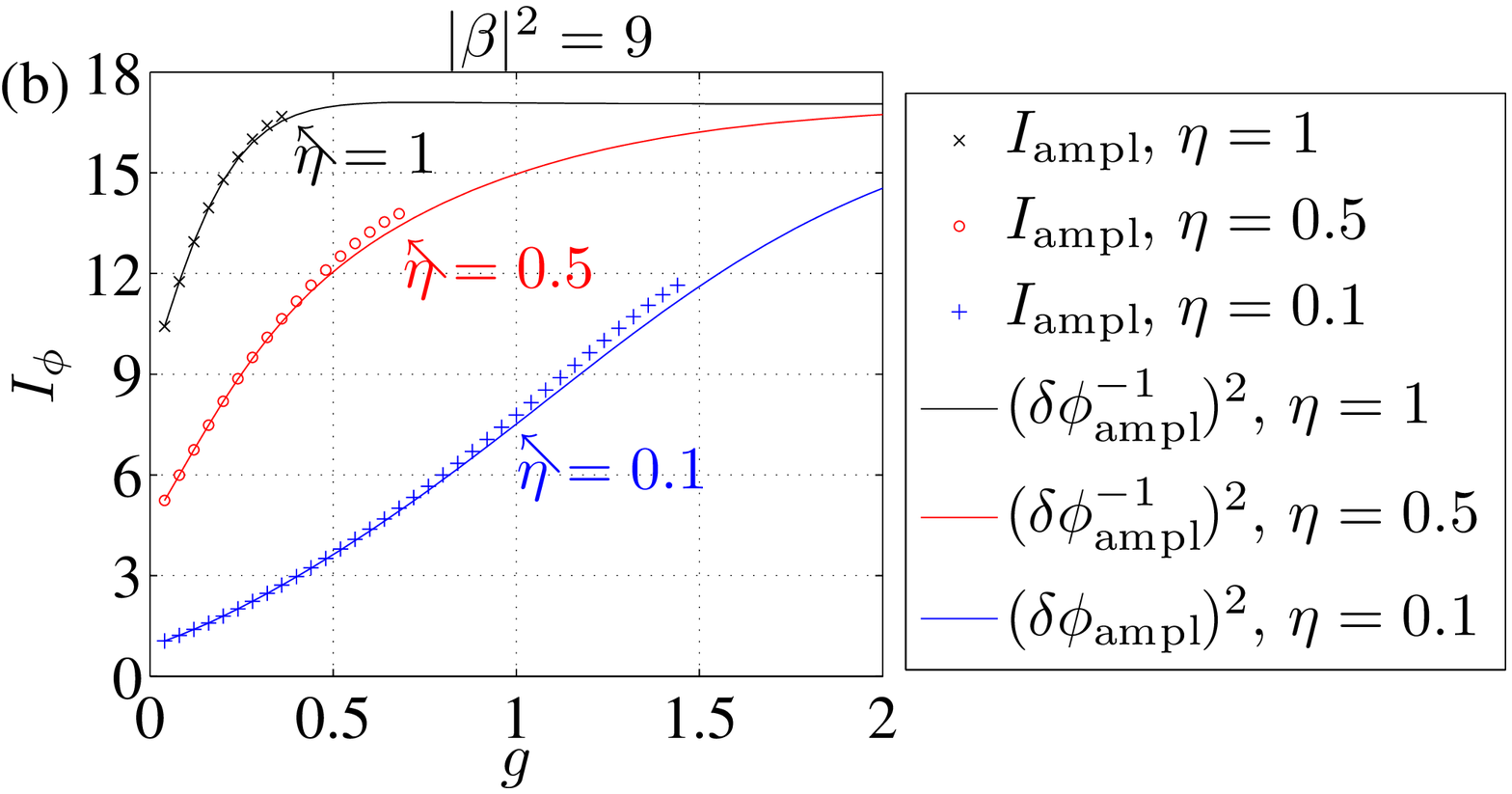}
\end{minipage}
\begin{minipage}{0.46\textwidth}
\includegraphics[width=0.99\textwidth]{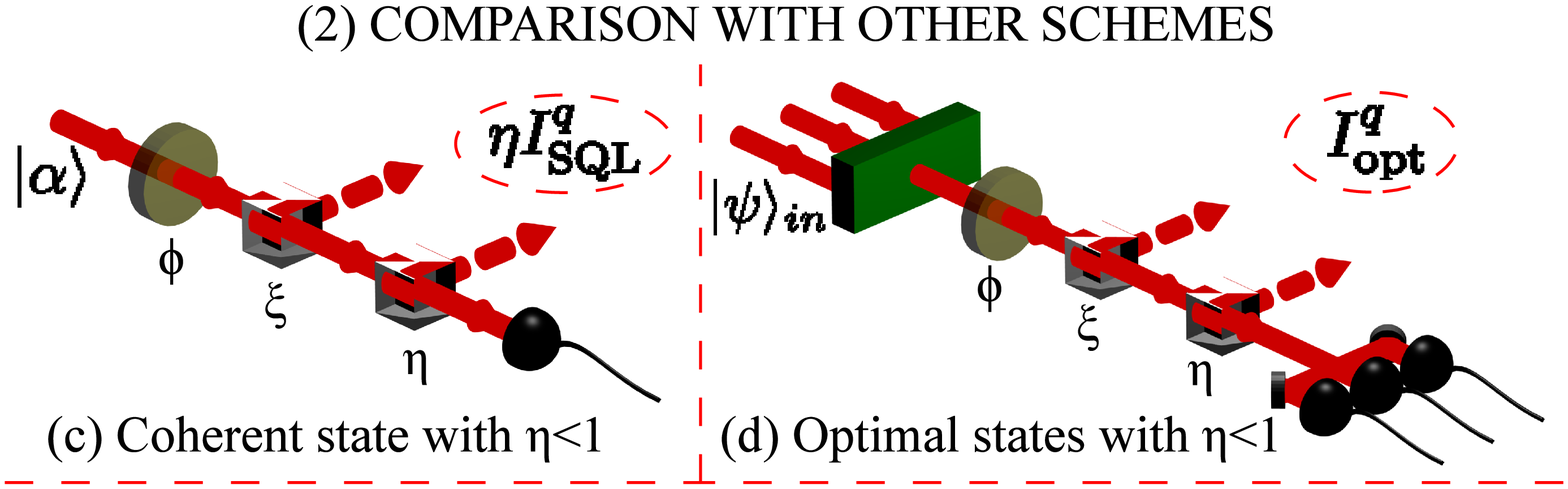}
\includegraphics[width=0.93\textwidth]{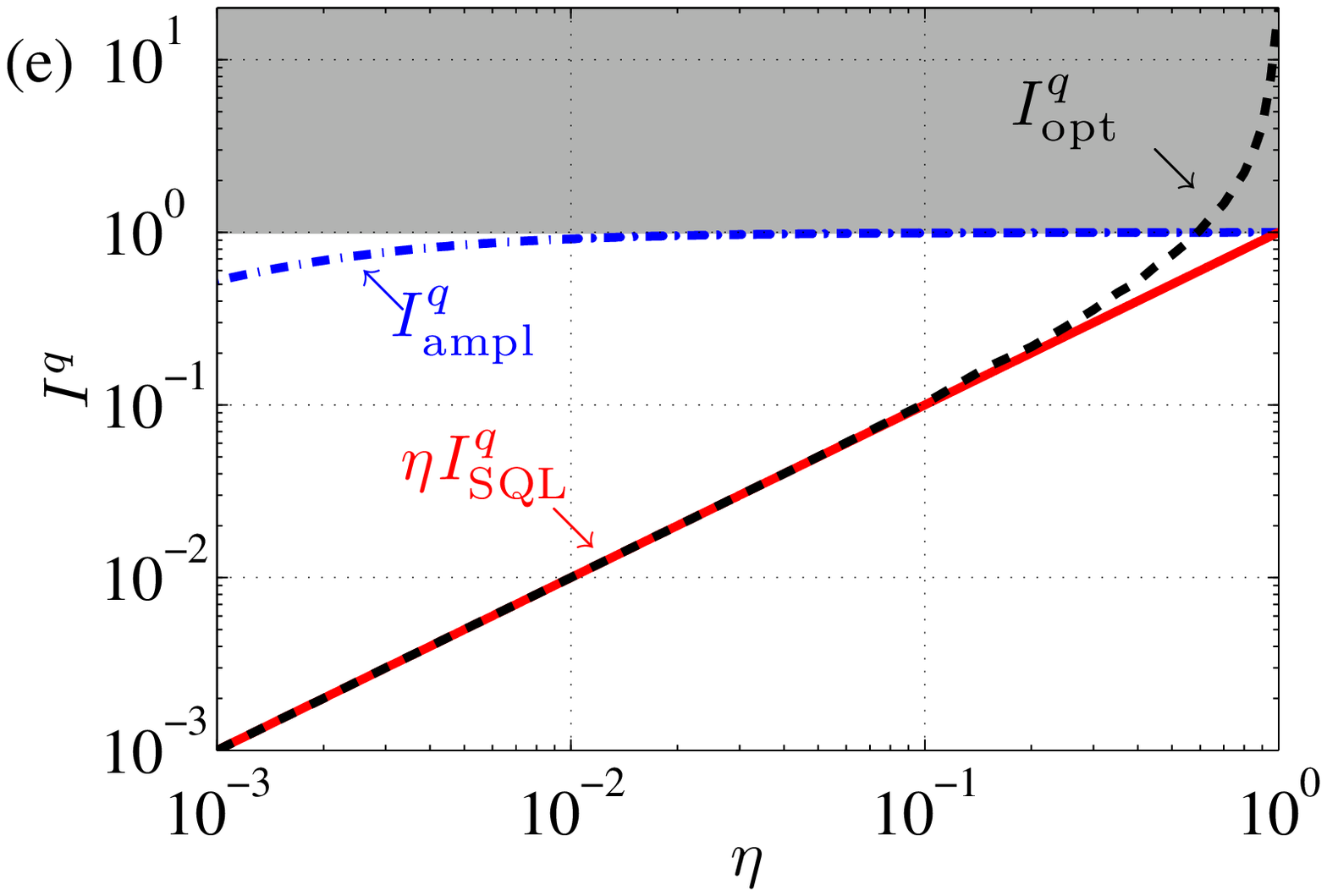}
\end{minipage}
\caption{(1) THEORETICAL ANALYSIS. (a)~Scheme of the
  amplifier based protocol. (b)~Comparison between the classical Fisher
  information $I_{\mathrm{ampl}}$ (points) and the sensitivity
  $(\delta \phi_{\mathrm{ampl}}^{-1})^{2}$ (lines) for $\vert \beta
  \vert^{2} = 9$. (2) COMPARISON WITH OTHER SCHEMES.  (c)~Conventional
  (unamplified) coherent-state interferometry with sample and
  detection loss $\mathcal{L}_{\xi}$ and $\mathcal{L}_{\eta}$ respectively: it can achieve the SQL
  bound connected to the quantum Fisher information (QFI)
  $\eta I^{q}_{\mathrm{SQL}}$. (d)~Interferometry based on the states that
  optimize the QFI in the presence of loss, proposed in \cite{Dorn09},
  with the corresponding QFI, $I^{q}_{\mathrm{opt}}$. (e)~Comparison between the
  QFI for the three strategies (a),(c) and (d), for $|\beta|^2=20$ and $g=3.5$, normalized with respect to
  $I^{q}_{\mathrm{SQL}}$. Blue dash-dotted line: QFI of our method $I^{q}_{\mathrm{ampl}}$.
  Red solid line: QFI of the coherent state phase estimation with loss
  of Fig.\ref{fig:1}c, $\eta I^{q}_{\mathrm{SQL}}$.  Green dashed line:
  QFI of the optimal strategy of Fig.\ref{fig:1}d, $I^{q}_{\mathrm{opt}}$ \cite{Dorn09}.
  }
\label{fig:1}
\end{figure*}
%
%

Our scheme employs a conventional interferometric phase
sensing stage that uses coherent-state probes. These are amplified
with an optical parametric amplifier (OPA) carrying the phase after the interaction with
the sample, but before the lossy detectors.  No post-selection is
employed to filter \cite{Giov11,post-sel-prl} the output signal. The
OPA (an optimal phase-covariant quantum cloning machine \cite{Naga07})
transfers the properties of the injected state into a field with a
larger number of particles, robust under losses and decoherence
\cite{Spag09}. Previous works addressed quantum signal amplification,
namely quadrature signal, in a lossy environment
adopting non-linear methods \cite{Leve93} and feedforward techniques \cite{Lam97}. 
At variance with these approaches our manuscript analyzes how the amplification of coherent 
states can be adopted for phase estimation purposes in a lossy environment. Specifically
by studying the quantum Fisher information problem, we show that, by adopting the 
amplification-based strategy, the extracted information can achieve the quantum 
Cram{\'e}r-Rao bound associated to the coherent probe state measured with a perfect
detection apparatus. Since the amplification acts after the
interaction of the probe state with the sample, our scheme is suitable
for the analysis of fragile samples, e.g.~optical microscopy of
biological cells \cite{Carl10}, or for single-photon interferometry
\cite{Vite10a} (where the small intensity of the probes achieved only
limited accuracy).

\textbf{\textit{Theory}} - The probe is a horizontally-polarized $(H)$ coherent state
$|\alpha \rangle_{H} \vert 0 \rangle_{V}$, with $\alpha = \vert \alpha
\vert e^{\imath \theta}$. The state is rotated in the $\vec{\pi}_{\pm}=
2^{-1/2}(\vec{\pi}_{H} \pm \vec{\pi}_{V})$ polarization basis, and the 
interaction with the sample induces a phase shift $\phi$ between the 
$\vec{\pi}_{\pm}$ polarization components: $U_{\phi}$. The sample
loss $\mathcal{L}_{\xi}$ reduces the state amplitude 
to $\beta=\sqrt{\xi}\alpha$. The maximum amount of information which can be extracted
on the coherent probe state is encoded in the corresponding QCR bound
$\delta \phi \geqslant (M I^{q}_{\mathrm{SQL}})^{-1/2}$, where $I^{q}_{\mathrm{SQL}}=
2|\beta|^2$. In the absence of amplification, the detection losses 
$\mathcal{L}_{\eta}$ would increase the QCR to $\delta 
\phi \geqslant (M \eta I^{q}_{\mathrm{SQL}})^{-1/2}$. To prevent this and
to attain the previous bound, we implemented the operations shown in
Fig.\ref{fig:1}a: a $\lambda/4$ wave-plate with optical axis at $45^{\circ}$ 
and the OPA, described by the unitary ${U}_{\mathrm{OPA}}=
\exp[ g (a_{H}^{\dag \; 2} - a_{V}^{\dag \; 2})/2 + \mathrm{h.c.}]$, where
$g=|g|e^{i\lambda}$ is the amplifier gain, and $a_H$ and $a_V$ are the
annihilation operators of the two polarization modes. After the action of detection losses $1-\eta$,
the state evolves into ${\rho}_{\phi}^{\beta,g,\eta}$. The quantum Fisher information
$I^{q}_{\mathrm{ampl}}$ of the amplification strategy, evaluated on the state ${\rho}_{\phi}^{\beta,g,\eta}$ and
quantifying the optimal performances of the scheme, reads
\begin{equation}
  \label{eq:QFI_amplified}
I^{q}_{\mathrm{ampl}}(\vert \beta \vert,g,\eta) = 2 \vert
\beta \vert^{2} \eta 
\frac{e^{2(g-g_{\mathrm{eff}})}}{\sqrt{1+4 \eta (1-\eta)
    \overline{n}}},
\end{equation}
where $g_{\mathrm{eff}}=1/4 \log[(\eta e^{2 g} + 1-\eta)/(\eta e^{-2
  g} + 1-\eta)]$, $\overline{n} = \sinh^{2} g$, and we maximized the $\phi$-dependent
quantum Fisher information by choosing $\phi=\pi/2-\lambda/2+\theta$ \cite{NotePhaseRef}. 
For $\overline{n} \gg (8 \eta)^{-1}$ and $\vert \beta \vert^{2} \gg 1/2$, 
we observe that $I^{q}_{\mathrm{ampl}}$ approaches the SQL limit $I_{\mathrm{SQL}}^{q}$
(dash-dotted line in Fig. \ref{fig:1}e). In other words, increasing the amplifier gain, the
effects of the detector loss can be asymptotically removed
\cite{Dall10}.

Achieving the accuracy associated with quantum Fisher bound $I^q_{\mathrm{ampl}}$ of 
(\ref{eq:QFI_amplified}) would need to use an optimal estimation strategy which is difficult to 
characterize~\cite{Pari09} and most likely challenging to implement. To experimentally test our proposal 
we decided hence to recover  $\phi$ by measuring (via lossy detectors) the photon number difference 
$D=n_H-n_V$ between the two modes on the output state ${\rho}_{\phi}^{\beta,g,\eta}$ 
after losses, with $n_x\equiv a_x^\dag a_x$. Even though in general this scheme fails to reach the 
accuracy bound of $I^q_{\mathrm{ampl}}$, in the limit of high gain $g$ and high amplitude $\beta$ it 
allows us to reach the value of $I^{q}_{\mathrm{SQL}}$ (and hence of  $I^q_{\mathrm{ampl}}$).
Indeed the resulting uncertainty can be evaluated \cite{Hels76} as $\delta \phi = \sigma(\langle D 
\rangle) \vert \frac{\partial \langle D \rangle}{\partial \phi}\vert^{-1}$, where $\<D\>$ is
the expectation value of $D$ on the output state. A calculation of the
estimation error $\delta\phi$ of the whole procedure shows that it depends on the value of the
phase $\phi$ to be estimated. The maximum sensitivity, that is,
the minimum uncertainty $\delta \phi_{\mathrm{ampl}}$,
is obtained for $\phi=\pi/2$ by setting $\lambda=2\theta$:
\begin{equation}
\delta \phi_{\mathrm{ampl}} = \frac{a^{1/2}(\overline{n},\eta)}{\vert \beta \vert^{2}
\sqrt{\eta}(1 + 2 \overline{n} + 2 \sqrt{\overline{n}(1+\overline{n})})},
\end{equation}
where $a(\overline{n},\eta) = 2 \overline{n} (1+\eta+2\eta \overline{n})+\vert \beta \vert^{2} 
\big[ 1 + 2 \overline{n} + \eta \overline{n} (6 + 8 \overline{n})\big]$.
It is then clear that for $\bar n\gg (2\eta)^{-1}$ and $|\beta|^2\gg1/2$
we have $\delta \phi_{\mathrm{ampl}} \simeq (2 |\beta|^{2})^{-1/2}$,
that is, the QCR bound of the state $|\Psi_\phi^\beta\>$ (before the amplification and the 
detector loss) can be attained by our detection strategy.
We also notice that the adopted data processing is optimal for a wide range
of parameters. This can be shown by evaluating the classical Fisher information
$I_{\mathrm{ampl}}$, which  represents the maximum amount of information that
can be extracted from the probe state using our choice of measurement,
optimizing over all possible data-processing. In the present strategy, the sensitivity 
$(\delta \phi_{\mathrm{ampl}}^{-1})^{2}$ closely tracks the
$I_{\mathrm{ampl}}$ both for small and intermediate values
of $\overline{n}$. Furthermore, the trend of the two curves suggest a close resemblance
also in the high photon number regime (see Fig.\ref{fig:1}b).

Because of the dependence of $\delta \phi$
on $\phi$, to achieve the minimum error $\delta \phi_{\mathrm{ampl}}$ an
adaptive strategy \cite{nagaoka-Barn00} is necessary. In the Supplementary
Material we show that it is sufficient to use a simple two-stage
strategy in which we first find a rough estimate of the phase
$\phi_{\mathrm{est}}$ employing conventional phase estimation methods,
and then we use it to tune the zero-reference so that our scheme
operates at its optimal working point detailed above. We also show
that the resources employed in the first stage of this adaptive
strategy are asymptotically negligible with respect to the resources
employed in the second high-resolution stage. 

\textbf{\textit{Efficiency of the phase estimation}} - 
We now compare our method to other strategies, using as a benchmark
the SQL $\delta \phi \geqslant(M I_{\mathrm{SQL}}^{q})^{-1/2}$, which
would be achieved by a probe coherent state with $\vert \beta
\vert^{2}$ average photons using lossless detectors.
Consider now the case with no amplification, where a
coherent state is subject to both the sample and detector loss
(Fig.\ref{fig:1}c). This is the strategy conventionally used in
interferometry \cite{nota_hom}. Our method clearly always outperforms it, see
the continuous line in Fig.\ref{fig:1}e. Furthermore, in a lossy scenario the present amplifier-based method
achieves better performances than any quantum strategy.
Recently, the optimal strategy in the presence of loss was derived
\cite{Dorn09} (Fig.\ref{fig:1}d). It employs the state that maximizes
the quantum Fisher information in lossy conditions. Of course, this
strategy cannot be beaten if one could access the optimal measurement
that attains the QCR bound. Even though elegant proof-of-principle
experiments exist \cite{Kacp10}, both this measurement and the
creation of these states without using post-selection are beyond the
reach of practical implementations for the foreseeable future,
especially for states with large average photon-numbers. In addition,
the form of these states strongly depends on the value of the loss
$\mathcal{L}_{\xi}$: it may be unknown and its experimental 
evaluation typically requires irradiating the sample, which removes the
advantage of using the optimal minimally-invasive states. In contrast,
the present amplifier-based protocol uses readily available input states and
detection strategies, and does not require a priori knowledge since
the choice of the coherent state is independent of the value of the
loss.  Since our method is devised especially to counter the detector
loss $\mathcal{L}_{\eta}$, we compare the performance of our states with the
optimal state calculated for the total amount of loss $\mathcal{L}_{\xi \eta}$,
showing that our method can achieve better performance for the
practically-relevant case of low values of $\eta$ (see dashed line in
Fig. \ref{fig:1}e), where the detection strategy is clearly not
optimized to achieve the QCR bound of the optimal states.

%
%
\begin{figure}[ht!]
\centering
\includegraphics[width=0.48\textwidth]{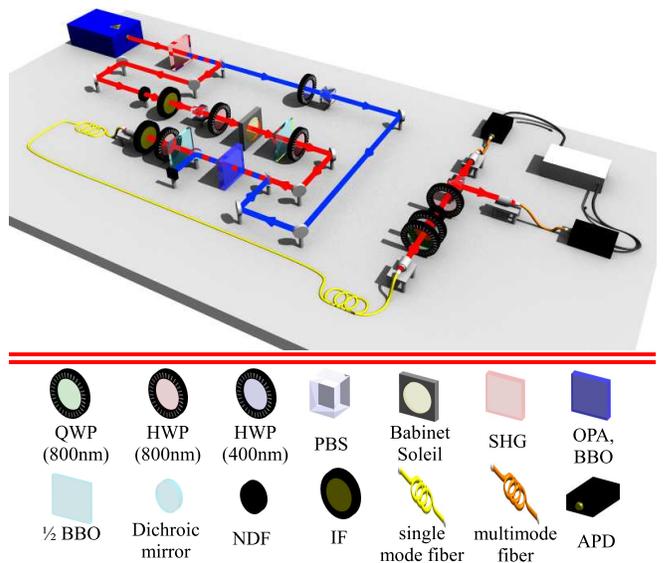}
\caption{Experimental setup for the practical implementation of the
  protocol. For details on the experimental setup refer to the description in
  the text and to the Supplementary Material.}
\label{fig:experiment}
\end{figure}
%
%

\textbf{\textit{Experimental Setup}} - We now describe the experimental
implementation in highly lossy conditions, showing that we can achieve
a significative phase-sensitivity enhancement with respect to the
coherent probe based strategy. The optical setup is reported in Fig.
\ref{fig:experiment}. To acquire the phase shift to be measured, the
probe coherent state is injected into the sample, which is simulated
by a Babinet-Soleil compensator that introduces a tunable phase shift $\phi$
between the ${H}$ and ${V}$ polarizations. Subsequently, the probe
state is superimposed spatially and temporally with a pump and
injected into the OPA. In this experimental realization the phases of
the pump and of the coherent state are not stabilized: this will
reduce the achievable enhancement by a fixed numerical factor of $4$.
Note that such condition corresponds to the absence of an external
phase reference.
In contrast to previous realizations of parametric amplification of
coherent states \cite{Zava04-Barb10} which focused on the
single-photon excitation regime, we could achieve a large value for
the nonlinear gain, up to $g=3.3$, corresponding to a number of
generated photons per mode $\overline{n} \sim 180$ in spontaneous
emission. In addition, our scheme is also able to exploit the
polarization degree of freedom.  After the amplification, the two
output orthogonal polarizations were spatially divided and detected by
two avalanche photodiodes. Their count rates are then subtracted to
obtain the value of $\<D\>$, and recorded as a function of the phase
$\phi$, introduced by the Babinet.

%
%
\begin{figure}[ht!]
  \centering
\includegraphics[width=0.235\textwidth]{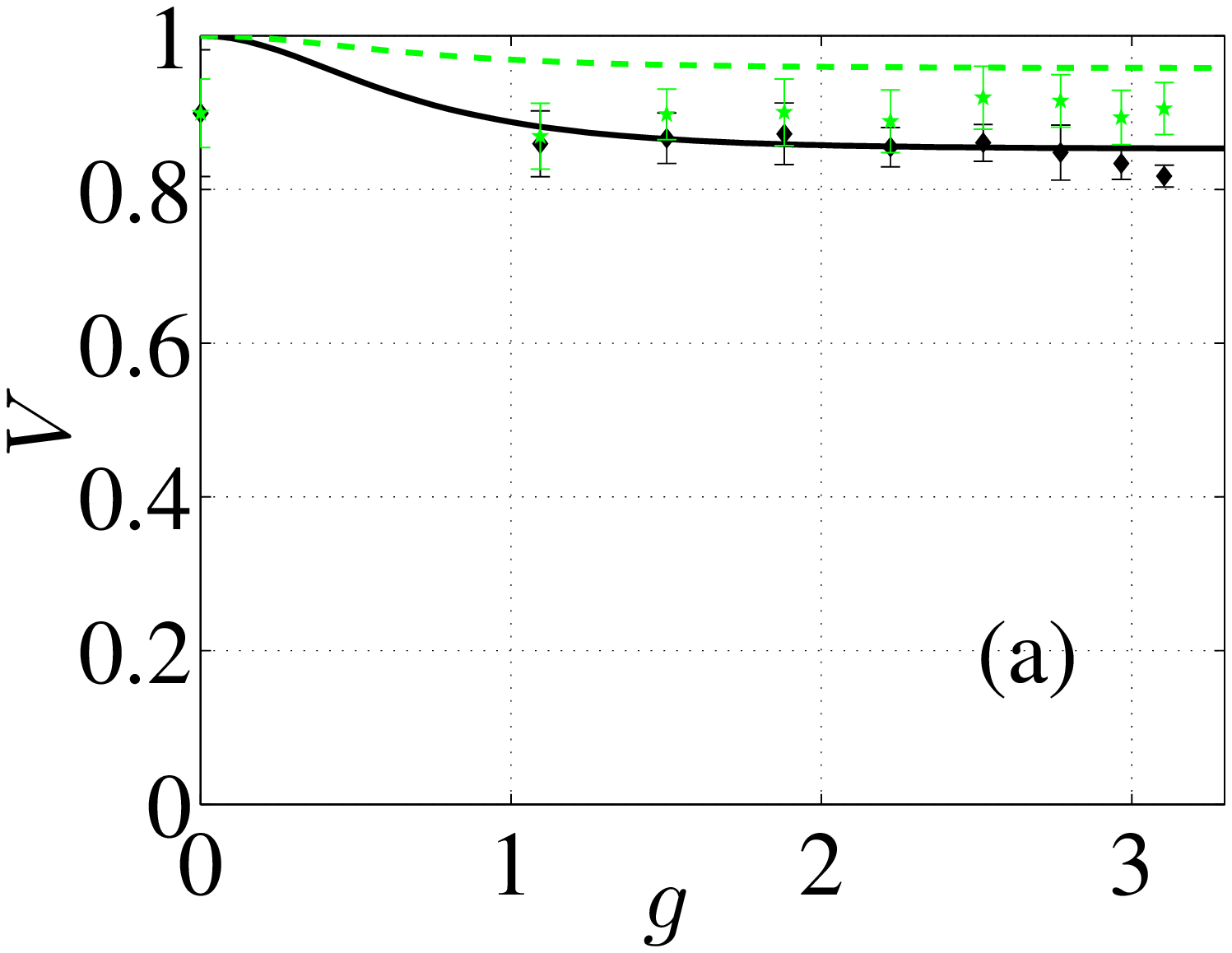}
\includegraphics[width=0.235\textwidth]{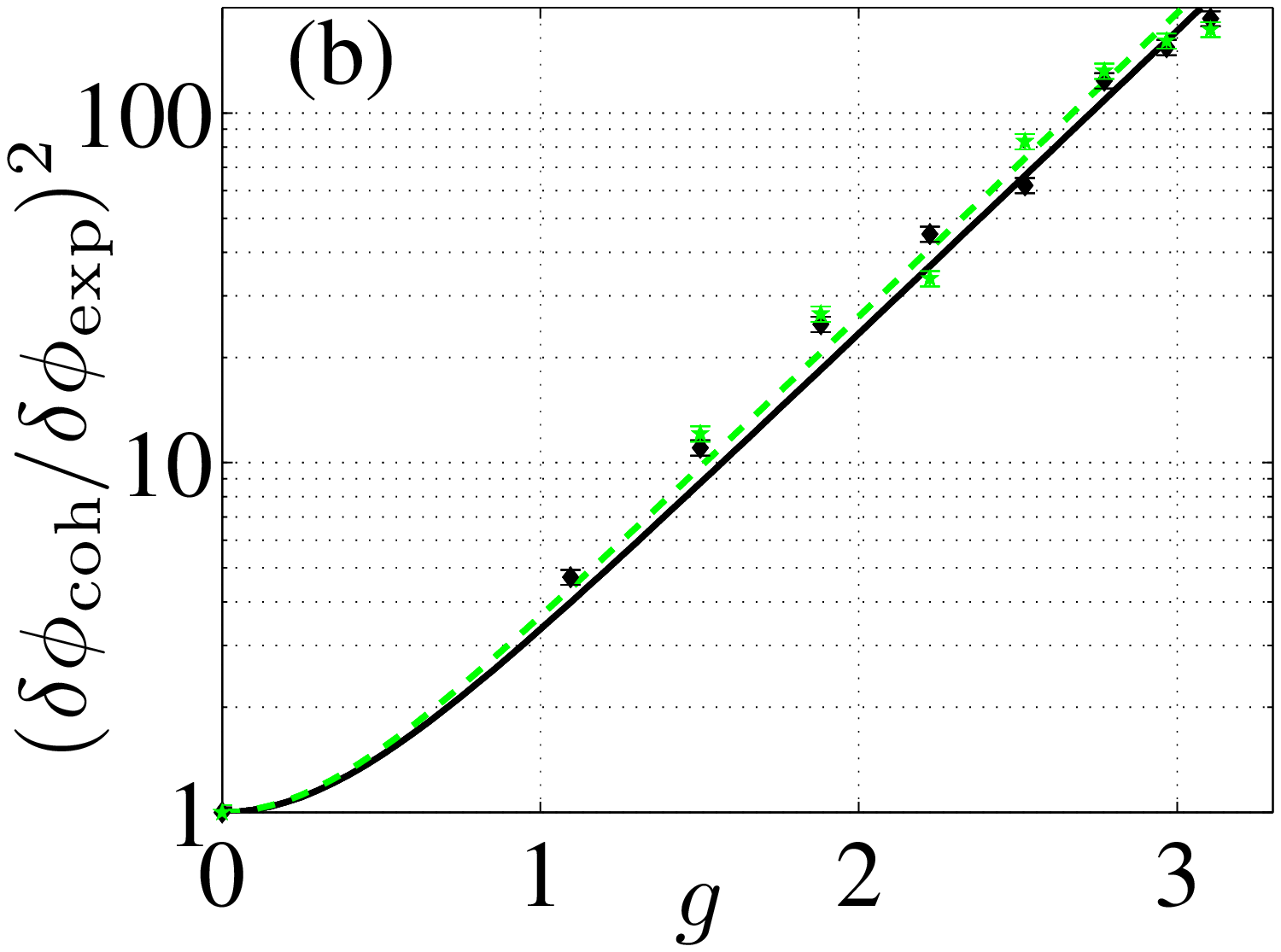}
\includegraphics[width=0.235\textwidth]{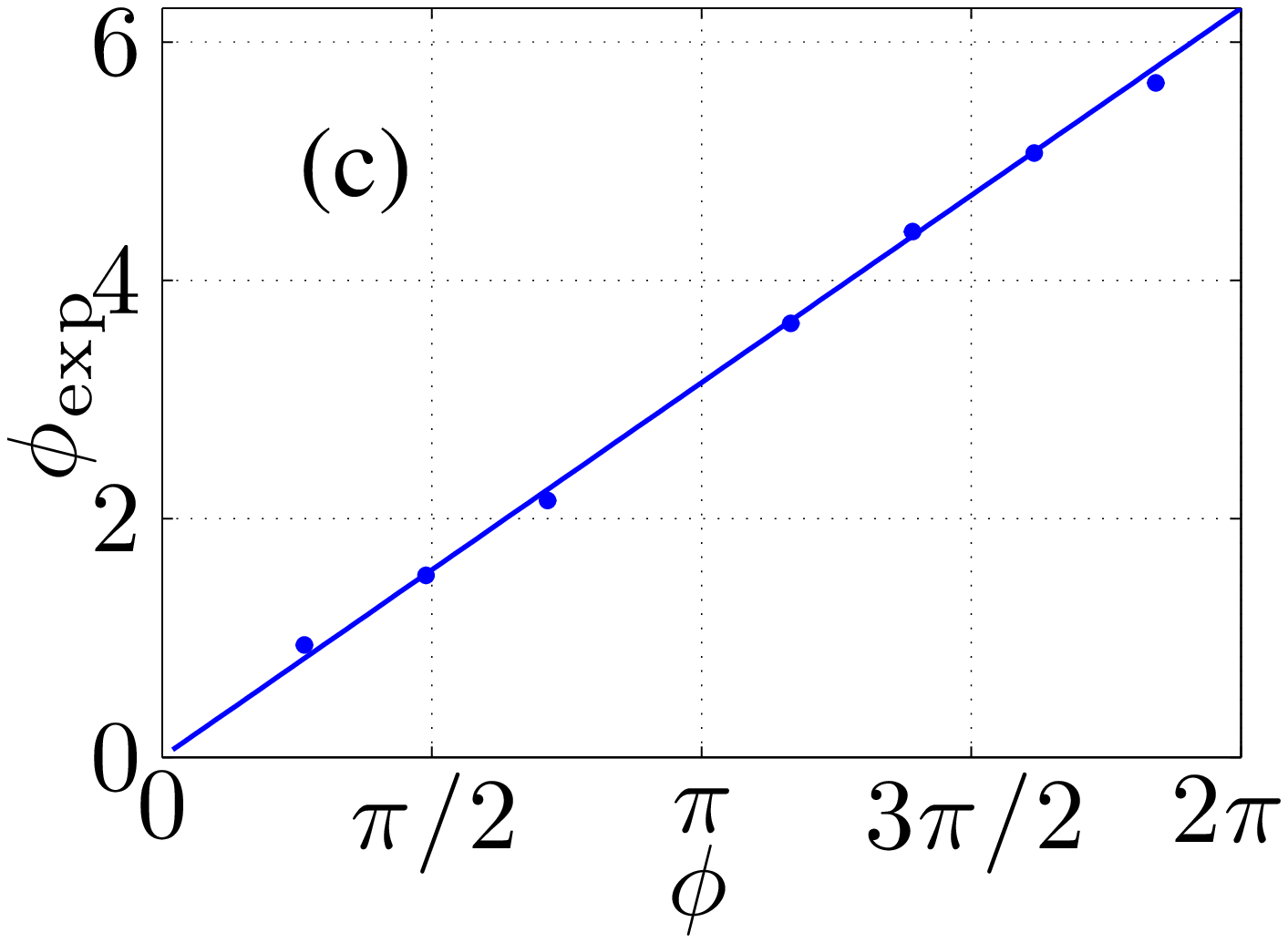}
\includegraphics[width=0.235\textwidth]{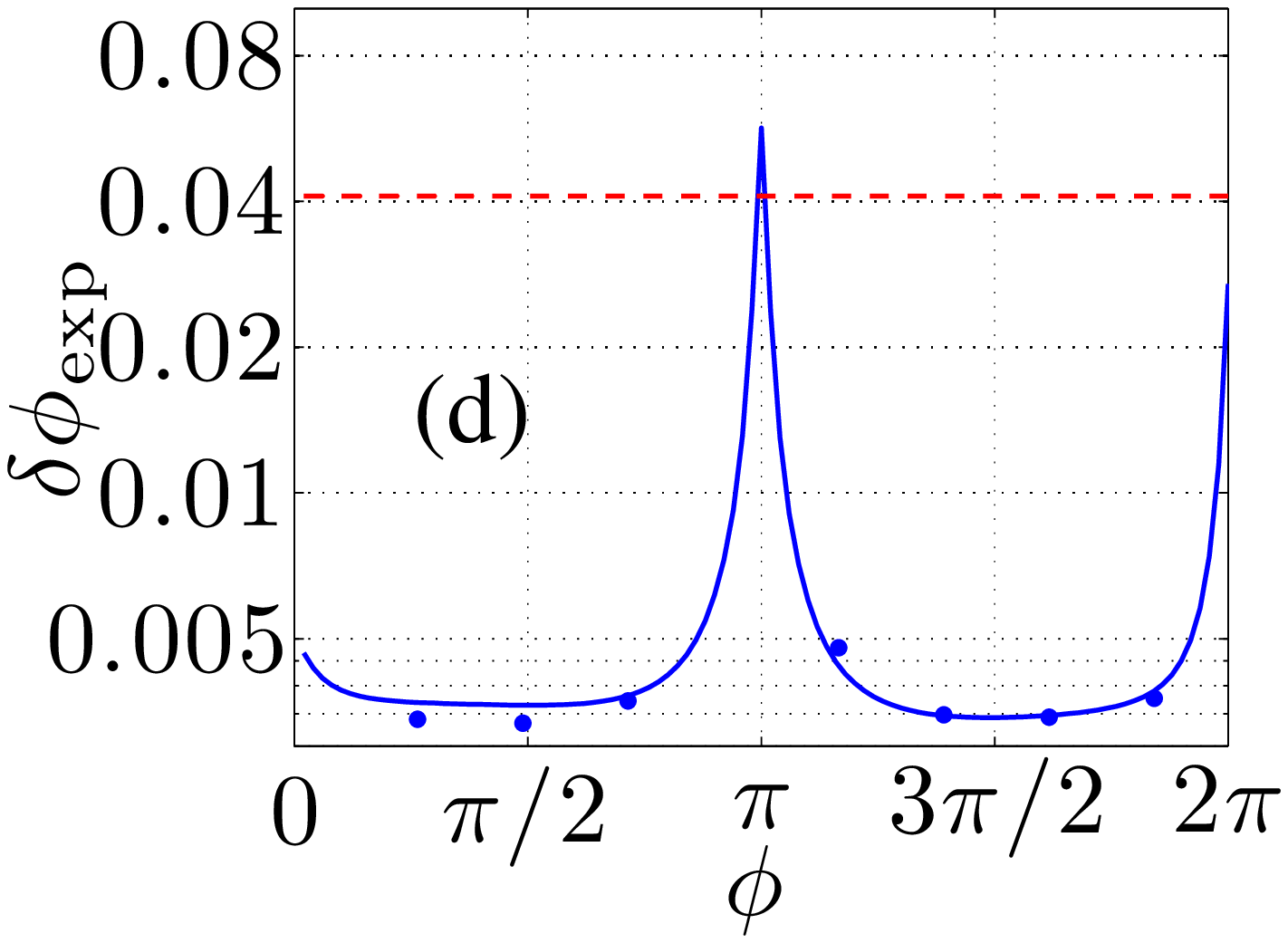}
\caption{Experimental results. (a)~Fringe pattern
  visibility and (b)~experimental enhancement $(\delta \phi_{\mathrm{coh}}/\delta 
  \phi_{\mathrm{exp}})^{2}$ evaluated at $\phi = \pi/2$ as a
  function of the nonlinear gain $g$ for $\vert \beta \vert^{2} \sim
  5.8$, $\eta \sim 1.46 \times 10^{-4}$ (experiment: black diamond
  points; theory: black solid line) and $\vert \beta \vert^{2} \sim
  22.8$, $\eta \sim 3.48 \times 10^{-5}$ (experiment: green star
  points; theory: green dashed line). (c)-(d) Experimental 
  results for the phase estimation experiment performed with the amplifier
  based strategy ($g=3.3$, $\vert \beta \vert^{2} \sim
  22.8$, $\eta \sim 3.48 \times 10^{-5}$) for different values of the phase. Estimated 
  values of the phase $\phi_{\mathrm{exp}}$ (c) and 
  corresponding error $\delta \phi_{\mathrm{exp}}$ (d). Points: experimental results. Blue solid 
  lines: theoretical prediction given respectively by the true value 
  of the phase $\phi$ (c) and by the classical Fisher 
  information (d). Red dashed line corresponds to the classical 
  Fisher information for the adopted coherent state without amplification.}
\label{fig:results}
\end{figure}
%
%

\textbf{\textit{Experimental phase estimation}} - The results 
of the experiment are reported in Fig. \ref{fig:results}.
An enhancement of $\sim 200$ in the counts rate for the former case is
observed without significantly affecting the visibility of the fringe
pattern (Fig. \ref{fig:results}a), leading to an increased phase
resolution. We measured the enhancement
$(\delta \phi_{\mathrm{coh}}/\delta \phi_{\mathrm{exp}})^{2}$ achievable
with our protocol $\delta \phi_{\mathrm{exp}}$ with respect to the conventional unamplified
interferometry $\delta \phi_{\mathrm{coh}}$, in the $\phi=\pi/2$ working point (see
Fig. \ref{fig:results}b). The quantity $(\delta \phi_{\mathrm{coh}}/\delta \phi_{\mathrm{exp}})^{2}$ 
represents the fraction of additional runs $\overline{M}$ of a coherent state phase 
estimation experiment in order to achieve the same performances of the amplifier-based
strategy, with the two protocols compared for the same values of 
$\vert \beta \vert^{2}$ and $\eta$. Our measurement shows a good agreement with
the theoretical predictions. A significant enhancement up to a value
of $(\delta \phi_{\mathrm{coh}}/\delta \phi_{\mathrm{exp}})^{2} = 186.3 \pm 9.3$ 
has been achieved.

We then performed a phase estimation experiment with the amplifier based
strategy for different values of the phase shift $\phi$. To this end, for each chosen value of the phase
we recorded the photon-counts in the two output detectors for $M_{\mathrm{exp}} = 7.5 \times
10^{5}$ subsequent pulses of the coherent state. Then, we adopted a Bayesian approach 
in order to obtain an estimate $\phi_{\mathrm{exp}}$ for the phase and to evaluate the associated
error $\delta \phi_{\mathrm{exp}}$. The results are reported in Figs. \ref{fig:results}c-d. We observe that
the estimated values of the phase $\phi_{\mathrm{exp}}$ are in good agreement with the corresponding 
true values $\phi$, and that the estimation process reaches the Cram{\'e}r-Rao bound. Furthermore, the
obtained results clearly outperforms the coherent state strategy when no amplification is performed (red dashed
line in Fig. \ref{fig:results}d.)

\textbf{\textit{Conclusions and perspectives}} - We discuss a strategy for phase estimation in the 
presence of noisy detectors that can reach the performance of a lossless probe. This 
approach involves coherent states as input signals, thus not requiring any a priori 
characterization of the amount of losses, and phase sensitive amplification after the 
interaction with the sample and before detection losses.  As a further perspective, our method
could be exploited with different classes of probe states, including quantum resources such as squeezing,
leading to sub-SQL phase estimation experiments in lossy conditions.

We acknowledge support by the FIRB ``Futuro in Ricerca'' Project HYTEQ, and Progetto
d'Ateneo of Sapienza Universit\`{a} di Roma. LM was supported by EU
through COQUIT, VG by MIUR through FIRB-IDEAS
Project RBID08B3FM.


\begin{thebibliography}{25}
\bibitem{NotePhaseRef} Note that in order to achieve the limit imposed by the QCR 
bound an external phase reference is required. In our case, such a reference is provided 
by the optical phase of the pump beam.
\bibitem{nota_hom} Other measurement strategies, involving for instance homodyne detection, could in principle reduce detection losses. However, homodyning requires extreme care in the mode-matching: this is typically extremely difficult and any failure in this will translate in reduced quantum efficiency. Similarly to photon counting, classical Fisher information corresponding to homodyning scales linearly with $\eta$: $I_{\mathrm{hom}}=2\eta|\beta|^{2}$.
\end{thebibliography}
\end{document}


\title{Supplementary Material: phase estimation via quantum interferometry for noisy
  detectors}

\author{Nicol\`{o} Spagnolo} \affiliation{Dipartimento di Fisica,
  Sapienza Universit\`{a} di Roma, piazzale Aldo Moro 5, I-00185 Roma,
  Italy} \affiliation{Consorzio Nazionale Interuniversitario per le
  Scienze Fisiche della Materia, piazzale Aldo Moro 5, I-00185 Roma,
  Italy} \author{Chiara Vitelli} \affiliation{Dipartimento di Fisica,
  Sapienza Universit\`{a} di Roma, piazzale Aldo Moro 5, I-00185 Roma,
  Italy} \author{Vito Giovanni Lucivero} \affiliation{Dipartimento di
  Fisica, Sapienza Universit\`{a} di Roma, piazzale Aldo Moro 5,
  I-00185 Roma, Italy} \author{Vittorio Giovannetti}
\affiliation{NEST, Scuola Normale Superiore and Istituto
  Nanoscienze-CNR, Piazza dei Cavalieri 7, I-56126 Pisa, Italy}
\author{Lorenzo Maccone} \affiliation{Dipartimento di Fisica ``A.
  Volta'', Universit\`{a} di Pavia, via A. Bassi 6, I-27100 Pavia,
  Italy} \author{Fabio Sciarrino} \email{fabio.sciarrino@uniroma1.it}
\homepage{http://quantumoptics.phys.uniroma1.it}
\affiliation{Dipartimento di Fisica, Sapienza Universit\`{a} di Roma,
  piazzale Aldo Moro 5, I-00185 Roma, Italy} \affiliation{Istituto
  Nazionale di Ottica, Consiglio Nazionale delle Ricerche (INO-CNR),
  largo Fermi 6, I-50125 Firenze, Italy}

\maketitle

In this supplementary material we elaborate on the material presented
in the main text, giving more details on the experimental
procedure and carefully deriving the formulas presented there. In
Sec.~\ref{s:experiment} we describe the experiment and the evolution
of the quantum state of the probe as it evolves through the apparatus.
In Sec.~\ref{s:state} we calculate the explicit
form of the output state of our apparatus. In Sec.~\ref{s:qfi} we
calculate the quantum Fisher information of the output state, and in
Sec.~\ref{s:cfi} the classical Fisher information that results from
fixing the detection scheme to the one we employ in the experiment.
In Sec.~\ref{s:sensi} we derive the phase error $\delta \phi_{\mathrm{ampl}}$ of our
apparatus, Eq.~(2) of the main text. In Sec.~\ref{s:adaptive} we give
the details of our simple two-stage adaptive scheme, showing how the
first stage (where a rough estimate of the phase $\phi$ is recovered)
can be neglected asymptotically, as it requires asymptotically
vanishing resources. We then simulate numerically the described two-step protocol.
Finally, in Sec.~\ref{s:exp} we give the details of the theoretical
model we employed to analyze the experimental data.

\section{Experimental setup}\label{s:experiment}

The probe is a horizontally ($H$) polarized electromagnetic field prepared in
the coherent state $\vert \alpha \rangle_{H} \vert 0 \rangle_{V}$ with
$\alpha = \vert \alpha \vert e^{\imath \theta}$. It is sent through an
interferometric setup to interact with the sample. The sample induces
a phase shift $\phi$ on the system and is characterized by a loss
$1-\xi$. The aim of our apparatus is to determine $\phi$, while
employing a low intensity signal. The phase shift is induced through a
unitary transformation of the type
\begin{eqnarray}
{U}_{\phi} = e^{-i({a}_-)^\dag a_- \phi},
\labell{unit}\;
\end{eqnarray}
where $a_-=(a_H-a_V)/\sqrt{2}$ is the annihilation operator connected
to the $-$ polarization. The loss is induced through a completely
positive map $\mathcal{L}_{\xi}$ of the form
\begin{eqnarray}
\begin{aligned}
  \mathcal{L}_{\xi}[\rho]=\sum_n A_n\rho (A_n)^\dag,
  \mbox{ with
  }A_n=\frac{(\xi^{-1}-1)^{n/2}}{\sqrt{n!}}a^n\xi^{\tfrac{a^\dag a}2},
\labell{loss}\;
\end{aligned}
\end{eqnarray}
where $\rho$ is an arbitrary state. Since the action of the phase
unitary $ U_\phi$ and of the loss $\mathcal{L}_\xi$ commute, we
can consider these two as independent processes that occur during the
interaction with the sample.  The action of the loss map on a coherent
state simply shifts its amplitude
$\mathcal{L}_\xi[|\alpha\>\<\alpha|]=|\sqrt{\xi}\alpha\>\<\sqrt{\xi}\alpha|$, without
changing the form of the state. Thus, our choice of coherent state
probes will not depend on the noise characteristics of the sample.
Consider first the unitary part of the interaction $ U_\phi$: the
state evolves as
\begin{equation}
\begin{aligned}
  \vert \Psi^{\alpha}_{\phi} \rangle &= {U}_{\phi} \vert \alpha
  \rangle_{H}  \vert 0 \rangle_{V}= \\
  &= \vert e^{- \imath \phi/2} \alpha \cos (\phi/2) \rangle_{H}
   \vert \imath e^{- \imath \phi/2} \alpha \sin (\phi/2)
  \rangle_{V}.
\end{aligned}
\end{equation}
Then, the action of the loss $\mathcal{L}_\xi$ reduces the amplitude of
the coherent states so that, after the interaction of the sample, the
probe has evolved to
\begin{equation}
  \vert \Psi^{\beta}_{\phi} \rangle = \vert e^{- \imath \phi/2}
  \beta \cos (\phi/2) \rangle_{H}  \vert \imath e^{- \imath
    \phi/2} \beta \sin (\phi/2) \rangle_{V},
\end{equation}
with $\beta=\sqrt{\xi}\alpha$. When this state is measured by a homodyne detection apparatus, the
error $\delta \phi$ on the phase $\phi$ reads $\delta \phi = (2 \vert \beta \vert^2 \eta)^{-1/2}$,
where $\eta$ is the overall detection efficiency which takes into account losses
and mode matching between the field and the local oscillator (spectral and spatial).
To overcome the limitation induced by $\eta$, we consider the following strategy.
Before the amplification, a relative
phase-shift of $\pi/2$ is inserted between the $H$ and the $V$
polarization components by means of a $\lambda/4$ birefringent
waveplate, leading to:
\begin{equation}
  \vert e^{- \imath \phi/2} \beta \cos (\phi/2)
  \rangle_{H}  \vert - e^{- \imath \phi/2} \beta \sin
  (\phi/2) \rangle_{V}.
\end{equation}
The resulting state is then injected in an optical parametric
amplifier (OPA).  The interaction Hamiltonian of the OPA is
\begin{equation}
{\mathcal{H}}_{OPA} = \imath \hbar \chi \left( {a}^{\dag}_+
  {a}^{\dag}_- \right) + \mathrm{H.c.}= \imath \hbar \chi \left(
  {{a}^{\dag}_{H}}^2-
  {{a}^{\dag}_{V}}^2 \right)/2 + \mathrm{H.c.}
\end{equation}
where $a_\pm=(a_H\pm a_V)/\sqrt{2}$, and $\chi$ is the parameter that
quantifies the strength of the interaction. It corresponds to a
unitary operation \begin{eqnarray}
{U}_{\mathrm{OPA}}=\exp[ g (a_{H}^{\dag \; 2}
- a_{V}^{\dag \; 2})/2 + \mathrm{h.c.}]
\labell{uopa}\;
\end{eqnarray}
where $g=|g|e^{i\lambda}=\chi t$ is the amplifier gain ($t$ being the
interaction time). Form the form of the unitary in \eqref{uopa}, it is
clear that the OPA is equivalent to two single-mode squeezers acting
independently on the modes $H$ and $V$ with opposite phases, namely
${U}_{\mathrm{OPA}}=S_H(-g)\otimes S_V(g)$, where $S_l(g)\equiv
\exp[ -g a_{l}^{\dag \; 2}/2+ \mathrm{h.c.}]$, $l=H,V$.

After the amplification, the state has evolved to
$|\Psi_{\phi}^{\beta,g}
\rangle={U}_{\mathrm{OPA}}|\Psi_\phi^\beta\rangle$.  Finally, it is
detected by lossy detectors, parametrized by a quantum efficiency
$\eta$. These are equivalent to perfect detectors that measure the
number of photons, preceded by a loss map $\mathcal{L}_\eta$
\cite{mkk}. The action of this map on the state
$|\Psi_{\phi}^{\beta,g} \rangle$ produces the mixed state
\begin{eqnarray}
{\rho}_{\phi}^{\beta,g,\eta}\equiv
\mathcal{L}_\eta[|\Psi_{\phi}^{\beta,g}
\rangle\langle\Psi_{\phi}^{\beta,g}|]
\labell{rhodef}\;.
\end{eqnarray}
The explicit form of this state will be calculated in
Sec.~\ref{s:state}.


The corresponding experimental setup for the present protocol is shown in 
Fig. 2 of the paper. The excitation source is a Ti:Sa laser system, consisting
in a Ti:Sa mode-locked Mira900, whose output beam is injected into
the Ti:Sa RegA9000 amplifier. The overall laser system can output a
$1.5$W beam at wavelength $\lambda=795$ nm.  In a first nonlinear
crystal, the output field is doubled in frequency through a second
harmonic generation (SHG) process to generate the experiment pump
beam at wavelength $\lambda_{p}=397.5$ nm of power $P=650$ W. The
remainder of the $795$ nm beam is then separated from the pump beam
through a dichroic mirror, and is prepared in the coherent state
$\vert \alpha \rangle_{+}$ by controlled attenuation, spectral
filtering (IF) and polarizing optics.  The coherent state probe then
acquires the phase shift by interacting with the sample (in our
case, a Babinet-Soleil compensator), and is then injected into the
OPA after the acquisition of the phase.

\section{State evolution}\label{s:state}
In this section we calculate the explicit form of the output state
$\rho_{\phi}^{\beta,g,\eta}$ of our scheme, by exploiting some
operatorial relations for Gaussian states. This will be useful to
evaluate the quantum and classical Fisher informations in the
following sections. The state impinging at the measurement stage after
detection losses can be written in the form:
\begin{equation}
\begin{aligned}
{\rho}_{\phi}^{\beta,g,\eta}&=\mathcal{L}_{\eta} \Big\{
{S}_{H}({g}_{H}) {S}_{V}({g}_{V}) \mathcal{L}_{\xi} \Big[
{D}_{H}(\alpha_{H}) {D}_{V}(\alpha_{V}) \vert 0 \rangle \langle 0 \vert
\\ &{D}_{H}^{\dag}(\alpha_{H}) {D}_{V}^{\dag}(\alpha_{V})
\Big]{S}^{\dag}_{H}({g}_{H}) {S}^{\dag}_{V}({g}_{V})\Big\}
\end{aligned}
\end{equation}
where ${D}_{l}(\alpha_{l})=\exp(\alpha_l{a_l}^\dag -\alpha_l^*a_l)$ is
the displacement operator such that $D(\alpha)|0\>=|\alpha\>$. The
action of the lossy channel $\xi$ and of the displacement operators
can be interchanged as
\begin{equation}
\begin{aligned}
  \mathcal{L}_{\xi} \Big[ {D}_{H}(\alpha_{H}) {D}_{V}(\alpha_{V})
  &\vert 0 \rangle \langle 0 \vert {D}_{H}^{\dag}(\alpha_{H})
  {D}_{V}^{\dag}(\alpha_{V}) \Big] = \\
  = {D}_{H}({\beta_{H}}) {D}_{V}({\beta_{V}}) &\vert 0 \rangle \langle
  0 \vert {D}_{H}^{\dag}({\beta}_{H}) {D}_{V}^{\dag}({\beta}_{V})
\end{aligned}
\end{equation}
where ${\beta}_{l} = \sqrt{\xi} \alpha_{l}$. The output state then reads:
\begin{equation}
\begin{aligned}
{\rho}_{\phi}^{\beta,g,\eta}&=\mathcal{L}_{\eta} \Big\{
{S}_{H}({g}_{H}) {S}_{V}({g}_{V}) {D}_{H}({\beta}_{H})
{D}_{V}({\beta}_{V}) \vert 0 \rangle \langle 0 \vert \\
&{D}_{H}^{\dag}({\beta}_{H})
{D}_{V}^{\dag}({\beta}_{V}){S}^{\dag}_{H}({g}_{H})
{S}^{\dag}_{V}({g}_{V})\Big\}
\end{aligned}
\end{equation}
The action of the squeezing operators and of the displacement
operators can be now inverted according to
\begin{eqnarray}
\label{eq:squeezing_displ_1}
{D}(\alpha) {S}({g}) &=& {S}({g}) {D}(\alpha_{+}) \\
\label{eq:squeezing_displ_2}
{S}({g}) {D}(\alpha) &=& {D}(\alpha_{-}) {S}({g})
\end{eqnarray}
where $\alpha_{\pm} \equiv \alpha \cosh g \pm \alpha^{\ast} e^{\imath
  \lambda} \sinh g$.  Using
Eqs. (\ref{eq:squeezing_displ_1}-\ref{eq:squeezing_displ_2}) we can
write
\begin{equation}
{S}_{l}({g}_{l}) {D}_{l}({\beta}_{l}) \vert 0 \rangle = {D}_{l}
(\gamma_{l}) {S}_{l}({g}_{l}) \vert 0 \rangle
\end{equation}
with $\gamma_{l} \equiv {\beta}_{l} \cosh g_{l} - {\beta}_{l}^{\ast}
e^{\imath \lambda_{l}} \sinh g_{l}$.  The output state can be then
written as
\begin{equation}
  \begin{aligned}
    {\rho}_{\phi}^{\beta,g,\eta}&=\mathcal{L}_{\eta} \Big\{
    {D}_{H}(\gamma_{H}) {D}_{V}(\gamma_{V}) {S}_{H}({g}_{H})
    {S}_{V}({g}_{V}) \vert 0 \rangle \langle 0 \vert \\
    &{S}^{\dag}_{H}({g}_{H}) {S}^{\dag}_{V}({g}_{V})
    {D}_{H}^{\dag}(\gamma_{H}) {D}_{V}^{\dag}(\gamma_{V})\Big\}
\end{aligned}
\end{equation}
By interchanging the action of the loss $\mathcal{L}_\eta$ and of the
displacement operators ${D}_{l}(\gamma_{l})$, we obtain
\begin{equation}
  \begin{aligned}
    {\rho}_{\phi}^{\beta,g,\eta}&= {D}_{H}(\tilde{\gamma}_{H})
    {D}_{V}(\tilde{\gamma}_{V})\mathcal{L}_{\eta} \Big\{ {S}_{H}({g}_{H})
    {S}_{V}({g}_{V}) \vert 0 \rangle \langle 0 \vert \\
    &{S}^{\dag}_{H}({g}_{H}) {S}^{\dag}_{V}({g}_{V}) \Big\}
    {D}_{H}^{\dag}(\tilde{\gamma}_{H}) {D}_{V}^{\dag}(\tilde{\gamma}_{V})
\end{aligned}
\end{equation}
where $\tilde{\gamma}_{l} = \sqrt{\eta} \gamma_{l}$. Finally, by
exploiting the identity (\ref{eq:squeezed_vacuum_losses}) of Appendix
\ref{sec:Appendix_CV_relations}, involving the action of
$\mathcal{L}_{\eta}$ on squeezed vacuum states, we can express the
output state after detection losses in the Gaussian form
\begin{equation}
\label{eq:output_state_simplified}
\begin{aligned}
  {\rho}_{\phi}^{\beta,g,\eta} &= {D}_{H}(\tilde{\gamma}_{H})
  {D}_{V}(\tilde{\gamma}_{V}) {S}_{H}({g}_{H}^{\mathrm{eff}})
  {S}_{V}({g}_{V}^{\mathrm{eff}}) \Big[
  {\rho}_{H}^{th}(N_{\mathrm{eff}}) \otimes \\
  &{\rho}_{V}^{th}(N_{\mathrm{eff}}) \Big]
  {S}^{\dag}_{H}({g}_{H}^{\mathrm{eff}})
  {S}^{\dag}_{V}({g}_{V}^{\mathrm{eff}})
  {D}_{H}^{\dag}(\tilde{\gamma}_{H})
  {D}_{V}^{\dag}(\tilde{\gamma}_{V})
\end{aligned}
\end{equation}
The expressions for ${g}_{l}^{\mathrm{eff}}$ and
$N_{l}^{\mathrm{eff}}$ are reported in 
Eqs. (\ref{eq:g_tau_eff_1}-\ref{eq:g_tau_eff_4}).

\subsection{Eigenvalues and Eigenvectors}
From Eq.~(\ref{eq:output_state_simplified}) one can calculate the
spectrum of eigenvalues and eigenvectors of
${\rho}_{\phi}^{\beta,g,\eta}$. As a first step, we observe that the
density matrix of the state takes the form of a separable state
${\rho}^{(H)}_{\phi} \otimes {\rho}^{(V)}_{\phi}$, where
\begin{equation} {\rho}^{(l)}_{\phi} = {D}_{l}(\tilde{\gamma}_{l})
  {S}_{l}({g}_{l}^{\mathrm{eff}})
  {\rho}_{l}^{th}(N_{l}^{\mathrm{eff}})
  {S}^{\dag}_{l}({g}_{l}^{\mathrm{eff}})
  {D}_{l}^{\dag}(\tilde{\gamma}_{l}),
\end{equation}
with $l=H,V$. Since the state for the two modes has the same Gaussian
form, the joint spectrum can be obtained by analyzing directly the
${\rho}^{(l)}_{\phi}$ single-mode state.  By expanding the density
matrix in the Fock basis we obtain:
\begin{equation}
\begin{aligned}
{\rho}^{(l)}_{\phi} &= \sum_{n=0}^{\infty}
\frac{(N_{l}^{\mathrm{eff}})^{n}}{(1+N_{l}^{\mathrm{eff}})^{n+1}}
{D}_{l}(\tilde{\gamma}_{l}) {S}_{l}({g}_{l}^{\mathrm{eff}}) \\ &\vert
n \rangle_{l} \langle n \vert {S}^{\dag}_{l}({g}_{l}^{\mathrm{eff}})
{D}_{l}^{\dag}(\tilde{\gamma}_{l})
\end{aligned}
\end{equation}
The eigenvalues and the eigenvectors of the state ${\rho}^{(l)}_{\phi}
= \sum_{n} \varrho_{n}^{(l)} \vert \psi_{n}^{(l)} \rangle_{l} \langle
\psi_{n}^{(l)} \vert$ are then respectively
\begin{eqnarray}
  \varrho^{(l)}_{n} &=&
  \frac{(N_{l}^{\mathrm{eff}})^{n}}{(1+N_{l}^{\mathrm{eff}})^{n+1}} \\
  \vert \psi_{n}^{(l)} \rangle_{l} &=& {D}_{l}(\tilde{\gamma}_{l})
  {S}_{l}({g}_{l}^{\mathrm{eff}}) \vert n \rangle_{l}
\end{eqnarray}
Finally, the eigenvalues and the eigenvectors of the joint two-modes
density matrix can be written as
\begin{eqnarray}
\label{eq:state_eigenvalues}
{\rho}_{\phi}^{\beta,g,\eta} &=& \sum_{m,n=0}^{\infty}
\varrho_{m,n} \vert \Psi_{m,n} \rangle_{HV} \langle \Psi_{m,n} \vert
\\
\varrho_{m,n} &=& \varrho_{m}^{(H)} \varrho_{n}^{(V)}\\
\label{eq:state_eigenvectors}
\vert \Psi_{m,n} \rangle_{HV} &=& \vert \psi_{m}^{(H)} \rangle_{H}
\otimes \vert \psi_{n}^{(V)} \rangle_{V}.
\end{eqnarray}

\section{Quantum Fisher Information}\label{s:qfi}
In this section we describe the calculation of the quantum Fisher
information (QFI) of the output state ${\rho}_{\phi}^{\beta,g,\eta}$
of our scheme.

The QFI for a generic mixed state ${\sigma} = \sum_{m} \sigma_{m}
\vert \zeta_{m} \rangle \langle \zeta_{m} \vert$, as reviewed in the
Appendix \ref{sec:Fisher_Information} in
Eq. (\ref{eq:Fisher_def_mixed}), can be evaluated as \cite{Pari09}:
\begin{equation}
  I^{q}_{\phi} = \sum_{p} \frac{(\partial_{\phi}
    \sigma_{p})^{2}}{\sigma_{p}} + 2 \sum_{n,m} \epsilon_{n,m} \vert
  \langle \zeta _{m}\vert \partial_{\phi} \zeta_{n} \rangle \vert^{2}
\end{equation}
Here $\sigma_{m}$ and $\vert \zeta_{m} \rangle$ are respectively the
eigenvalues and the eigenvectors of the density matrix, and
$\epsilon_{n,m} =
(\sigma_{n}-\sigma_{m})^{2}/(\sigma_{n}+\sigma_{m})$. In the case of
the output density matrix ${\rho}_{\phi}^{\beta,g,\eta}$ of the
amplifier-based protocol the eigenvalues and the eigenvectors are
parametrized by the indices $(n,m)$, and the QFI is
\begin{equation}
\label{eq:Fisher_definition_OPA}
\begin{aligned}
I^{q}(\alpha,\xi,&\{g_{l}\},\{\lambda_{l}\},\eta) = \sum_{p,q=0}^{\infty}
\frac{(\partial_{\phi} \varrho_{p,q})^{2}}{\varrho_{p,q}}+ \\
&+2 \sum_{i,j,m,n=0}^{\infty} \epsilon_{i,j,m,n} \vert \langle
\Psi_{i,j} \vert \partial_{\phi} \Psi_{m,n} \rangle \vert^{2}
\end{aligned}
\end{equation}
where
\begin{equation}
\epsilon_{i,j,m,n} =
\frac{(\varrho_{i,j}-\varrho_{m,n})^{2}}{\varrho_{i,j}+\varrho_{m,n}}.
\end{equation}
We observe that, for the density matrix
${\rho}_{\phi}^{\beta,g,\eta}$, the eigenvalues $\varrho_{m,n}$
(\ref{eq:state_eigenvalues}) are independent on the phase $\phi$, and
hence the first term in Eq. (\ref{eq:Fisher_definition_OPA}) vanishes.
In order to calculate the second term, it is necessary to evaluate the
following quantity: $\vert \langle \Psi_{i,j} \vert \partial_{\phi}
\Psi_{m,n} \rangle \vert^{2}$. Such term can be written as
\begin{equation}
\label{eq:scalar_intermediate}
\begin{aligned}
& \langle \Psi_{i,j} \vert \partial_{\phi} \Psi_{m,n} \rangle =
\langle \Psi_{i,j} \vert \partial_{\phi} \big(  \vert \psi_{m}^{(1)}
\rangle_{1} \otimes \vert \psi_{n}^{(2)} \rangle_{2} \big) = \\
&= \langle \Psi_{i,j} \vert \big(  \vert \partial_{\phi}
\psi_{m}^{(1)} \rangle_{1} \otimes \vert \psi_{n}^{(2)} \rangle_{2} +
\vert \psi_{m}^{(1)} \rangle_{1} \otimes \vert \partial_{\phi}
\psi_{n}^{(2)} \rangle_{2} \big) = \\ &= \,_{1}\langle \psi^{(1)}_{i}
\vert \partial_{\phi} \psi_{m}^{(1)} \rangle_{1} \delta_{j,n} +
\delta_{i,m} \,_{2}\langle \partial_{\phi} \psi_{i}^{(2)} \vert
\psi_{m}^{(2)} \rangle_{2}
\end{aligned}
\end{equation}
Since the eigenvectors for the two-modes present an analogous form, it
is necessary to evaluate only the term $_{l}\langle \psi^{(l)}_{i}
\vert \partial_{\phi} \psi_{m}^{(l)} \rangle_{l}$.
Let us focus on the $\vert \partial_{\phi} \psi_{m}^{(l)} \rangle_{l}$
state vector. Since the dependence on $\phi$ of the state is included
only in the displacement operator ${D}_{l}(\tilde{\gamma}_{l})$, we
can write:
\begin{equation}
\label{eq:partial_psi_m}
\vert \partial_{\phi} \psi_{m}^{(l)} \rangle_{l} =
\big[\partial_{\phi} {D}_{l}(\tilde{\gamma}_{l})\big]
{S}_{l}({g}_{l}^{\mathrm{eff}}) \vert m \rangle_{l}
\end{equation}
The latter can be evaluated by differentiating the displacement
operator written in normally-ordered form:
\begin{equation}
\partial_{\phi} \big[ {D}_{l}(\tilde{\gamma}_{l}) \big] =
\partial_{\phi} \big[ e^{-\frac{1}{2} \vert \tilde{\gamma}_{l}
  \vert^{2}} e^{\tilde{\gamma}_{l}{a}_{l}^{\dag}}
e^{-\tilde{\gamma}_{l}^{\ast} {a}_{l}} \big]
\end{equation}
By differentiating the three exponential with respect to $\phi$, and
by exploiting the following commutation relation:
\begin{equation}
\big[ {a}_{l}, e^{\tilde{\gamma}_{l} {a}_{l}^{\dag}} \big] =
\tilde{\gamma}_{l} e^{\tilde{\gamma}_{l} {a}_{l}^{\dag}}
\end{equation}
the derivative of ${D}_{l}(\tilde{\gamma}_{l})$ reads: 
\begin{equation}
\partial_{\phi} \big[ {D}_{l}(\tilde{\gamma}_{l}) \big] = \big[
C^{(l)}_{\alpha,\xi,g_{l},\lambda_{l},\eta,\phi} +
F^{(l)}_{\alpha,\xi,g_{l},\lambda_{l},\eta,\phi}({a}_{1},{a}_{l}^{\dag})
\big] {D}_{l}(\tilde{\gamma}_{l}).
\end{equation}
The scalar $C^{(l)}_{\alpha,\xi,g_{l},\lambda_{l},\eta,\phi}$ and the
operator
$F^{(l)}_{\alpha,\xi,g_{l},\lambda_{l},\eta,\phi}({a}_{l},{a}_{l}^{\dag})$
are respectively:
\begin{eqnarray}
C^{(l)}_{\alpha,\xi,g_{l},\lambda_{l},\eta,\phi} &=& \frac{1}{2} \big[
\tilde{\gamma}_{l} (\partial_{\phi} \tilde{\gamma}_{l}^{\ast}) -
(\partial_{\phi} \tilde{\gamma}_{l}) \tilde{\gamma}_{l}^{\ast} \big]
\\
F^{(l)}_{\alpha,\xi,g_{l},\lambda_{l},\eta,\phi}({a}_{l},{a}_{l}^{\dag})
&=& (\partial_{\phi} \tilde{\gamma}_{l}) {a}_{l}^{\dag} -
(\partial_{\phi} \tilde{\gamma}_{l}^{\ast}) {a}_{l}
\end{eqnarray}
By replacing the latter expressions in Eq. (\ref{eq:partial_psi_m}),
the scalar product $_{l}\langle \psi_{i}^{(l)} \vert \partial_{\phi}
\psi_{m}^{(l)} \rangle_{l}$ can be evaluated as:
\begin{equation}
\begin{aligned}
_{l}\langle \psi^{(l)}_{i} \vert \partial_{\phi} \psi_{m}^{(l)}
\rangle_{l} &= \,_{l}\langle i \vert
{S}^{\dag}_{l}({g}_{l}^{\mathrm{eff}})
{D}_{l}^{\dag}(\tilde{\gamma}_{l}) \big[
C^{(l)}_{\alpha,\xi,g_{l},\lambda_{l},\eta,\phi} + \\
&+
F^{(l)}_{\alpha,\xi,g_{l},\lambda_{l},\eta,\phi}({a}_{l},{a}_{l}^{\dag}
) \big] {D}_{l}(\tilde{\gamma}_{l}) {S}_{l}({g}_{l}^{\mathrm{eff}})
\vert m \rangle_{l}.
\end{aligned}
\end{equation}
Such average value can be evaluated by exploiting the operatorial
identities
\begin{eqnarray}
\label{eq:squeezing_evolution_1}
{S}^{\dag}({g}) {a} {S}({g}) &=& {a} \cosh g - {a}^{\dag} e^{\imath \lambda} \sinh g \\
\label{eq:squeezing_evolution_2}
{S}^{\dag}({g}) {a}^{\dag} {S}({g}) &=& {a}^{\dag} \cosh g - {a} e^{-\imath \lambda} \sinh g\\
\label{eq:displacement_evolution_1}
{D}^{\dag}(\alpha) {a} {D}(\alpha) &=& {a} + \alpha\\
\label{eq:displacement_evolution_2}
{D}^{\dag}(\alpha) {a}^{\dag} {D}(\alpha) &=& {a}^{\dag} + \alpha^{\ast}.
\end{eqnarray}
 We obtain
\begin{equation}
\begin{aligned}
\label{eq:Fisher_scalar_intermediate}
_{l}\langle \psi^{(l)}_{i} \vert \partial_{\phi} &\psi_{m}^{(l)}
\rangle_{l} = \delta_{i,m}
A^{(l)}_{\alpha,\xi,g_{l},\lambda_{l},\eta,\phi} - \delta_{i,m-1}
\sqrt{m} \times \\ &\times
B^{(l)\,\ast}_{\alpha,\xi,g_{l},\lambda_{l},\eta,\phi} +
\delta_{i,m+1} \sqrt{m+1}
B^{(l)}_{\alpha,\xi,g_{l},\lambda_{l},\eta,\phi}
\end{aligned}
\end{equation}
where the $A^{(l)}_{\alpha,\xi,g_{l},\lambda_{l},\eta,\phi}$ and
$B^{(l)}_{\alpha,\xi,g_{l},\lambda_{l},\eta,\phi}$ quantities are
defined as
\begin{eqnarray}
A^{(l)}_{\alpha,\xi,g_{l},\lambda_{l},\eta,\phi} &=& \frac{1}{2} \big[
(\partial_{\phi} \tilde{\gamma}_{l}) \tilde{\gamma}_{l}^{\ast} -
\tilde{\gamma}_{l} (\partial_{\phi} \tilde{\gamma}_{l}^{\ast}) \big]\\
B^{(l)}_{\alpha,\xi,g_{l},\lambda_{l},\eta,\phi} &=& \cosh
g_{l}^{\mathrm{eff}} (\partial_{\phi} \tilde{\gamma}_{l}) - e^{\imath
  \lambda_{l}} \sinh g_{l}^{\mathrm{eff}} (\partial_{\phi}
\tilde{\gamma}_{l}^{\ast})\nonumber\\&&
\end{eqnarray}
Note that the $\epsilon_{i,j,m,n}$ coefficients present the following
symmetries,
\begin{eqnarray}
\epsilon_{m,n,m,n} &=& 0\\
\epsilon_{i,j,m,n} &=& \epsilon_{m,j,i,n}\\
\epsilon_{i,j,m,n} &=& \epsilon_{i,n,m,j}
\end{eqnarray}
By inserting
Eqs. (\ref{eq:scalar_intermediate})-(\ref{eq:Fisher_scalar_intermediate})
in Eq. (\ref{eq:Fisher_definition_OPA}) and by exploiting the symmetries
of the $\epsilon_{i,j,m,n}$ coefficients we obtain
\begin{equation}
\label{eq:Fisher_general}
\begin{aligned}
I^{q}(\alpha,\xi,&\{g_{l}\},\{\lambda_{l}\},\eta) = 4
\sum_{m,n=0}^{\infty} \big[ \vert
B^{(1)}_{\alpha,\xi,g_{l},\lambda_{l},\eta} \vert^{2} (m+1)  \\
& \times \epsilon_{m+1,n,m,n} +\vert
B^{(2)}_{\alpha,\xi,g_{l},\lambda_{l},\eta} \vert^{2} (n+1)
\epsilon_{m,n,m,n+1}\big]
\end{aligned}
\end{equation}

\begin{figure}[ht!]
\centering
\includegraphics[width=0.35\textwidth]{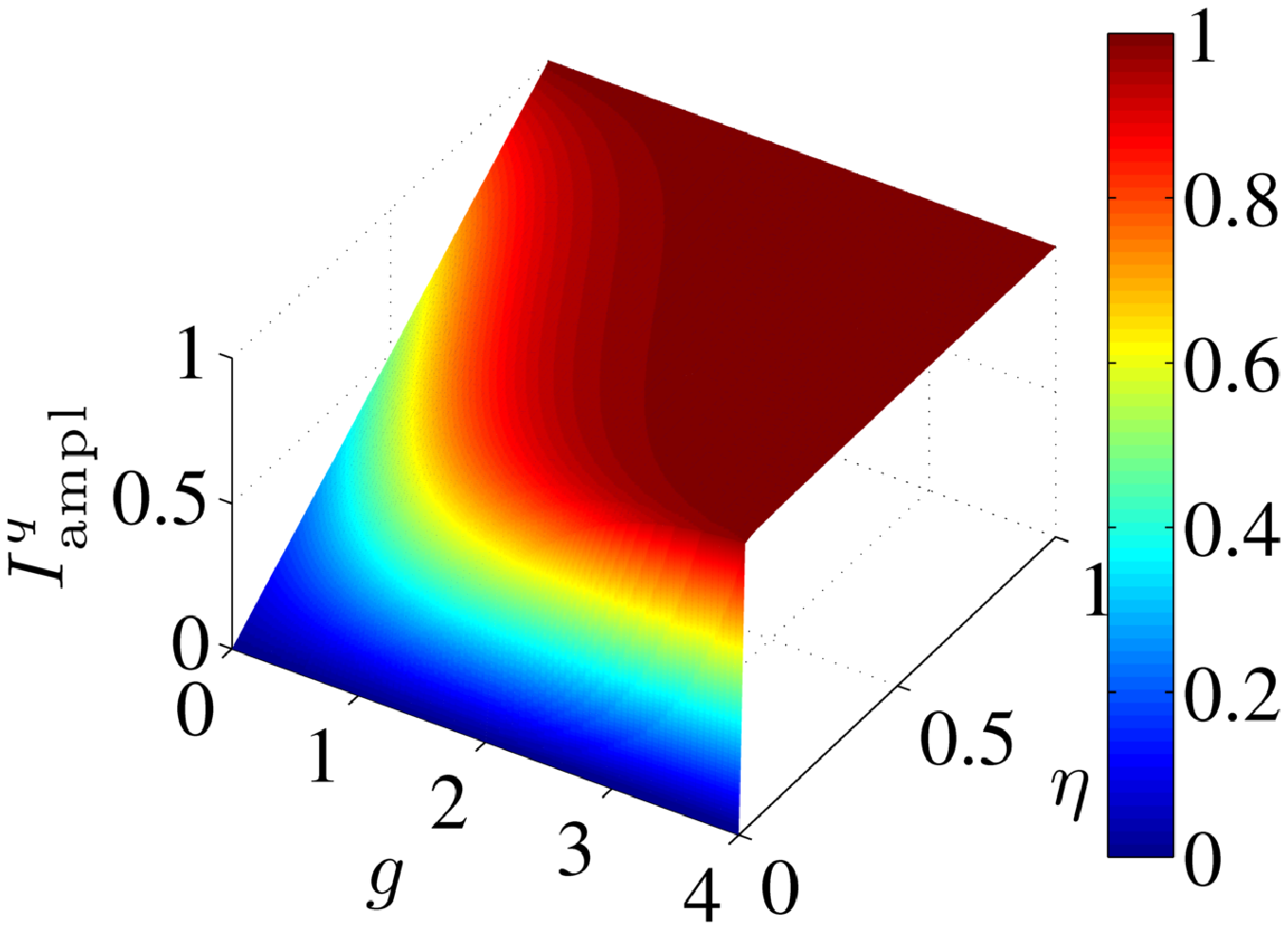}
\caption{Plot of $I^{q}_{\mathrm{ampl}}$ as a function of the nonlinear gain $g$ of the amplifier and of the 
  detection efficiency $\eta$, with $\vert \beta \vert^{2} = 20$, normalized with respect to $I^{q}_{\mathrm{SQL}}$}
\label{fig:figS1}
\end{figure}

The QFI $I^{q}_{\mathrm{ampl}}(\alpha,\theta,\phi,\xi,g,\lambda,\eta)$ of the
scheme is obtained by replacing ${g}_{H} \rightarrow - {g}$ and
${g}_{V} \rightarrow - {g}$. This choice of the parameters is
equivalent to the case described in the main paper (with ${g}_{H}
\rightarrow - {g}$, ${g}_{V} \rightarrow {g}$ and the additional
$\pi/2$ phase shift in the probe state) leading to the same expression
for the QFI.  We finally obtain
\begin{equation}
\begin{aligned}
&I^{q}(\alpha,\theta,\phi,\xi,g,\lambda,\eta) = \frac{2
  \vert \alpha \vert^{2} \xi \eta}{\sqrt{1+4 \eta (1-\eta) \sinh^{2}
    g}} \times \\
&\big\{ \cosh[2(g-g_{\mathrm{eff}})]-\cos(\lambda+2\phi-2\theta)
\sinh[2(g-g_{\mathrm{eff}})]\big\}\label{QFI}
\end{aligned}
\end{equation}
The optimal condition corresponds to the case
$\cos(\lambda+2\phi-2\theta)=-1$, where the QFI is
\begin{equation}
  I^{q}_{\mathrm{ampl}}(\alpha,\xi,g,\eta) = 2 \vert \alpha \vert^{2} \xi
  \eta \frac{e^{2(g-g_{\mathrm{eff}})}}{\sqrt{1+4 \eta (1-\eta)
      \sinh^{2} g}}.
\end{equation}
In Fig. \ref{fig:figS1} we report the trend of $I_{\mathrm{ampl}}^{q}$ normalized with
respect to the SQL $I_{\mathrm{SQL}}^{q}$, and we observe that for  $\bar n\gg (8\eta)^{-1}$ 
and $|\beta|^2\gg1/2$ we have $I^{q}_{\mathrm{ampl}} \rightarrow I_{\mathrm{SQL}}^{q}$.
Again, the dependence of the QFI $I^{q}$ of \eqref{QFI} on the parameter
$\phi$ to be estimated implies that to achieve its maximum
$I^{q}_{\mathrm{ampl}}$, an adaptive strategy (see Sec.~\ref{s:adaptive})
is necessary.

\section{Classical Fisher information for the photon-counting
  measurement}\label{s:cfi}
\begin{figure}[ht!]
\centering
\includegraphics[width=0.25\textwidth]{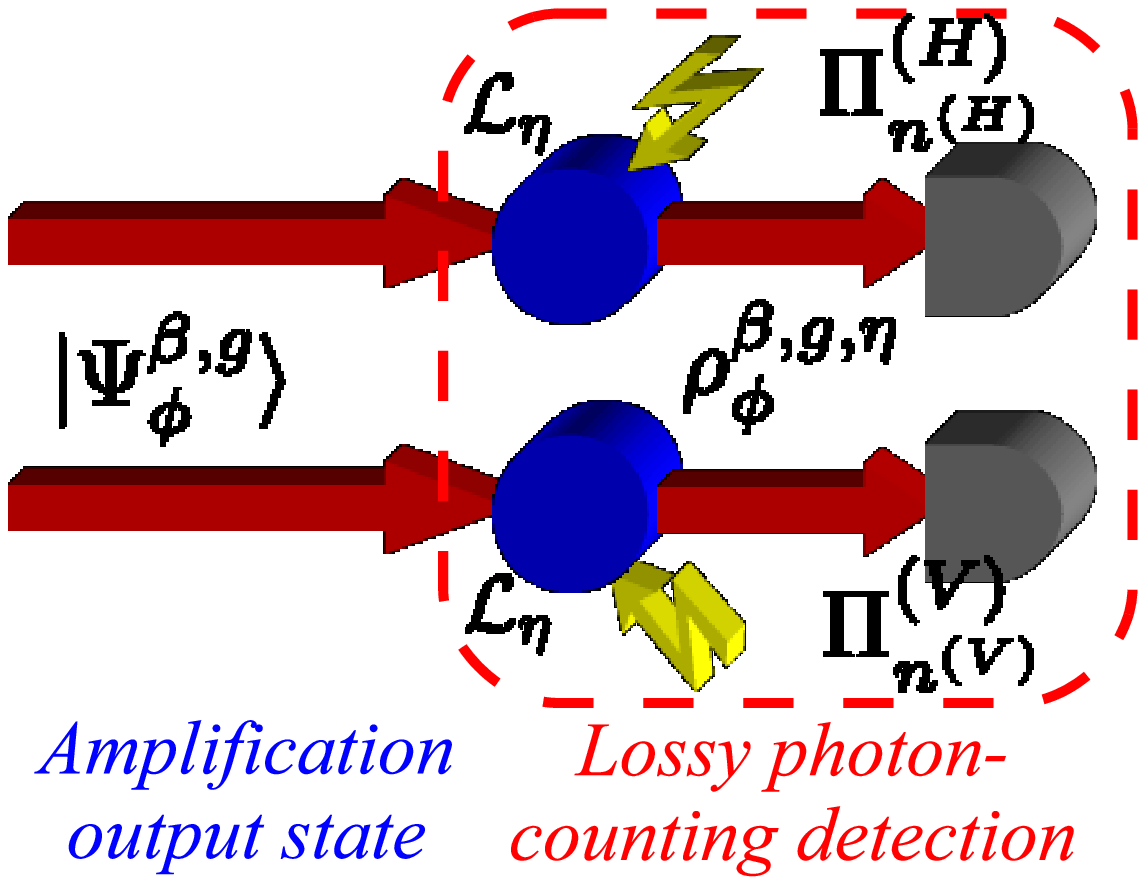}
\caption{Scheme of the different steps of the amplifier-based phase
  estimation protocol when photon-counting measurements are
  performed.}
\label{fig:schema_photon_counting}
\end{figure}

In this section we describe the calculation for the classical Fisher
information associated with our scheme when photon-counting
measurements are performed [Fig.\ref{fig:schema_photon_counting}].
The output state of the protocol is described by the density matrix
${\rho}_{\phi}^{\beta,g,\eta}$, while the measurement operators that
describe photon-counting detectors are the projectors over Fock
states
\begin{equation}
{\Pi}_{n^{(H)},n^{(V)}} = {\Pi}_{n^{(H)}}^{(H)} \otimes {\Pi}_{n^{(V)}}^{(V)}
\end{equation}
where ${\Pi}_{n^{(l)}}^{(l)} = \vert n^{(l)} \rangle_{l} \,
_{l}\langle n^{(l)} \vert$, with $l=H,V$ labeling the optical
mode. The probability distribution of the measurement outcomes can be
evaluated as
\begin{equation}
  p(n^{(H)},n^{(V)} \vert \phi) = \mathrm{Tr}
  [{\rho}_{\phi}^{\beta,g,\eta} {\Pi}_{n^{(H)},n^{(V)}}]
\end{equation}
The classical Fisher information associated to the probability
distributions of the measurement outcomes is given by the following
expression \cite{Pari09}:
\begin{equation}
I_{\phi}= \sum_{n,m=0}^{\infty} \frac{[\partial_{\phi}
  p(n^{(H)},n^{(V)} \vert \phi)]^{2}}{p(n^{(H)},n^{(V)} \vert \phi)}
\end{equation}
For the amplifier-based protocol, the probability distribution
$p(n^{(H)},n^{(V)} \vert \phi)$ can be separated in two independent
single-mode contributions as
\begin{equation}
p(n^{(H)},n^{(V)} \vert \phi) = \prod_{l=H,V} p(n^{(l)} \vert \phi)
\end{equation}
Here, ${\rho}^{(l)}$ are the single-mode density matrices for modes
$l=H,V$ and
\begin{equation}
p(n^{(l)} \vert \phi) = \mathrm{Tr}[{\rho}^{(l)} {\Pi}^{(l)}_{n^{(l)}}]
\end{equation}
In this case, the classical Fisher information can be separated in two
single-mode contributions
\begin{equation}
\label{eq:CFI_def_1}
I_{\phi} = \sum_{l=H,V} I_{\phi}^{(l)}
\end{equation}
where
\begin{equation}
\label{eq:CFI_def_2}
I_{\phi}^{(l)} = \sum_{n=0}^{\infty} \frac{[\partial_{\phi} p(n^{(l)}
  \vert \phi)]^{2}}{p(n^{(l)} \vert \phi)}
\end{equation}

\subsection{Photon-number distribution of the amplified coherent
  states}
We begin by calculating the photon-number distribution of the
amplified coherent states. The density matrix of the output state
before the measurement stage is given by
\begin{equation}
  \begin{aligned}
    {\rho}_{\phi}^{\beta,g,\eta} &= {D}_{H}(\tilde{\gamma}_{H})
    {D}_{V}(\tilde{\gamma}_{V}) {S}_{H}({g}_{H}^{\mathrm{eff}})
    {S}_{V}({g}_{V}^{\mathrm{eff}}) \Big[
    {\rho}_{H}^{th}(N^{\mathrm{eff}}_{H}) \otimes \\
    &{\rho}_{V}^{th}(N^{\mathrm{eff}}_{V}) \Big]
    {S}^{\dag}_{H}({g}_{H}^{\mathrm{eff}})
    {S}^{\dag}_{V}({g}_{V}^{\mathrm{eff}})
    {D}_{H}^{\dag}(\tilde{\gamma}_{H})
    {D}_{V}^{\dag}(\tilde{\gamma}_{V})
\end{aligned}
\end{equation}
to evaluate the photon-number distribution, we exploit the following
identity between the elements of the density matrix expressed in the
Fock basis ${\rho} = \sum_{n,m=0}^{\infty} \rho_{n,m} \vert n \rangle
\langle m \vert$ and the Wigner function of a general single-mode
state ${\rho}$,
\begin{equation}
\rho_{n,m} = \pi \int_{-\infty}^{\infty} \int_{-\infty}^{\infty} dx dp
W_{\rho}(x,p) W_{n,m}(x,p)
\end{equation}
where $W_{n,m}(x,p)$ is the Wigner function associated to the operator
$\vert n \rangle \langle m \vert$. Here, the $({x},{p})$ operators are
defined according to $\Delta^{2}x \Delta^{2}p \geq 1/16$. The
corresponding photon-number distribution can be recovered from the
diagonal elements $\rho_{n,n}$, by exploiting the expression of the
Wigner function of a Fock state:
\begin{equation}
W_{n,n}(x,p) = \frac{2}{\pi} (-1)^{n} L_{n}[4(x^{2}+p^{2})]
e^{-2(x^{2}+p^{2})}
\end{equation}
Since the density matrix of the state ${\rho}_{\phi}^{\beta,g,\eta} =
{\rho}_{\phi}^{(H)} \otimes {\rho}_{\phi}^{(V)}$ is separable between
the two modes, we can evaluate the distributions for the two
components ${\rho}_{\phi}^{(l)}$ separately. The first step is the
evaluation of the Wigner function for the single-mode density matrix:
\begin{equation}
{\rho}_{\phi}^{(l)} = {D}_{l}(\tilde{\gamma}_{l})
{S}_{l}({g}_{l}^{\mathrm{eff}}) {\rho}_{l}^{th}(N_{l}^{\mathrm{eff}})
{S}^{\dag}_{l}({g}_{l}^{\mathrm{eff}})
{D}_{l}^{\dag}(\tilde{\gamma}_{l})
\end{equation}
The Wigner function for this state takes the following Gaussian
form
\begin{equation}
\begin{aligned}
W_{\rho^{(l)}}(x_{l},p_{l}) &= \frac{2}{\pi} \frac{1}{1 + 2
  N^{\mathrm{eff}}_{l}} e^{-\frac{2}{1 + 2 N^{\mathrm{eff}}_{l}}[2
  (x_{l}-x_{l}^{0}) (p_{l}-p_{l}^{0}) \sigma^{xp}_{l}]} \\
&\times e^{-\frac{2}{1 + 2 N^{\mathrm{eff}}_{l}}
  [(x_{l}-x_{l}^{0})^{2} \sigma^{xx}_{l} + (p_{l}-p_{l}^{0})^{2}
  \sigma^{pp}_{l}]}
\end{aligned}
\end{equation}
where the first order and the second order moments are, respectively
\begin{eqnarray}
x_{l}^{0} &=& \mathrm{Re}[\tilde{\gamma_{l}}]\\
p_{l}^{0} &=& \mathrm{Im}[\tilde{\gamma_{l}}]
\end{eqnarray}
and
\begin{eqnarray}
\sigma_{l}^{xx} &=& \cosh (2 g_{l}^{\mathrm{eff}}) + \cos \lambda_{l}
\sinh (2 g_{l}^{\mathrm{eff}})\\
\sigma_{l}^{pp} &=& \cosh (2 g_{l}^{\mathrm{eff}}) - \cos \lambda_{l}
\sinh (2 g_{l}^{\mathrm{eff}}) \\
\sigma_{l}^{xp} &=& \sin \lambda_{l} \sinh (2 g_{l}^{\mathrm{eff}})
\end{eqnarray}
Here, $g^{\mathrm{eff}}_{l}$ and $\lambda_{l}$ are respectively the
absolute values and the phase of the squeezing parameters
${g}^{\mathrm{eff}}_{l}$. We can now proceed with the calculation of
the single-mode photon-number distribution $p(n^{(l)} \vert \phi)$,
which can be evaluated from the integral
\begin{equation}
\label{eq:integral_wigner}
p(n^{(l)} \vert \phi) = \pi \int_{-\infty}^{\infty} \int_{-\infty}^{\infty} 
dx_{l} dp_{l}W_{\rho^{(l)}}(x_{l},p_{l}) W_{n,m}(x_{l},p_{l})
\end{equation}
We first begin by performing the following rotation on the quadrature 
variables $(x_{l},p_{l}) \rightarrow (x'_{l},p'_{l})$ of the $W_{\rho^{(l)}}(x,p)$ 
function:
\begin{eqnarray}
x'_{l} &=& x_{l} \cos \psi_{l} + p_{l} \sin \psi_{l}\\
p'_{l} &=& - x_{l} \sin \psi_{l} + p_{l} \cos \psi_{l}\\
x^{\prime \; 0}_{l} &=& x_{l}^{\prime \; 0} \cos \psi_{l} +
p_{l}^{\prime \; 0} \sin \psi_{l}\\
p^{\prime \; 0}_{l} &=& - x_{l}^{\prime \; 0} \sin \psi_{l} +
p_{l}^{\prime \; 0} \cos \psi_{l}
\end{eqnarray}
where $\psi_{l} = \lambda_{l}/2$. The Wigner function in this rotated
quadrature set is
\begin{equation}
\begin{aligned}
W_{\rho^{(l)}}(x'_{l},p'_{l}) &= \frac{2}{\pi} \frac{1}{1 + 2 N^{\mathrm{eff}}_{l}} 
e^{-\frac{2}{1 + 2 N^{\mathrm{eff}}_{l}} [(x'_{l}-x_{l}^{\prime \, 0})^{2} e^{2 
g_{l}^{\mathrm{eff}}}]} \\ &\times e^{-\frac{2}{1 + 2 N^{\mathrm{eff}}_{l}} 
[(p'_{l}-p_{l}^{\prime \, 0})^{2} e^{- 2 g_{l}^{\mathrm{eff}}}]}
\end{aligned}
\end{equation}
The same rotation is performed on the $W_{n,n}(x_{l},p_{l})$, which
presents radial symmetry and hence its form is not affected by the
rotation according to
\begin{equation}
W_{n,n}(x'_{l},p'_{l}) = \frac{2}{\pi} (-1)^{n}
L_{n}\{4[(x'_{l})^{2}+(p'_{l})^{2}]\}
e^{-2[(x'_{l})^{2}+(p'_{l})^{2}]}
\end{equation}
We can then proceed with the evaluation of the integral
(\ref{eq:integral_wigner}). By performing the basis rotation
$(x_{l},p_{l}) \rightarrow (x'_{l},p'_{l})$ in the integration
variable we obtain
\begin{equation}
p(n^{(l)} \vert \phi) = \pi \int_{-\infty}^{\infty}
\int_{-\infty}^{\infty} dx'_{l} dp'_{l}W_{\rho^{(l)}}(x'_{l},p'_{l})
W_{n,m}(x'_{l},p'_{l})
\end{equation}
By expanding the Laguerre polynomials of the $W_{n,n}(x'_{l},p'_{l})$
function we obtain
\begin{equation}
\begin{aligned}
p(n^{(l)} \vert \phi) &= \frac{4 (-1)^{n}}{\pi (1+2
  N_{l}^{\mathrm{eff}})} \sum_{k=0}^{n} \frac{n!}{k!(n-k)!}
\sum_{j=0}^{k} \frac{(-4)^{k}}{k!} \begin{pmatrix} k \\ j
\end{pmatrix} \\
& \times \int_{-\infty}^{\infty} \int_{-\infty}^{\infty} dx'_{l}
dp'_{l} (x'_{l})^{j} (p'_{l})^{k-j}
e^{-2[(x'_{l})^{2}+(p'_{l})^{2}]}\\
& \times e^{-\frac{2}{1 + 2 N^{\mathrm{eff}}_{l}}
  [(x'_{l}-x_{l}^{\prime \, 0})^{2} e^{2 g_{l}^{\mathrm{eff}}} +
  (p'_{l}-p_{l}^{\prime \, 0})^{2} e^{- 2 g_{l}^{\mathrm{eff}}}]}
\end{aligned}
\end{equation}
The integrals in $d x'_{l}$ and $d p'_{l}$ can be evaluated
separately. We now define the following auxiliary functions
\begin{eqnarray}
\tilde{A}_{x_{l}} &=& 1+\frac{e^{-2 g^{\mathrm{eff}}_{l}}}{1+2 N^{\mathrm{eff}}_{l}}\\
\tilde{B}_{x_{l}} &=& \frac{x^{\prime \, 0}_{l} e^{-2 g^{\mathrm{eff}}_{l}}}{1 + 
2 N^{\mathrm{eff}}_{l} + e^{-2 g^{\mathrm{eff}}_{l}}} \\
\tilde{C}_{x_{l}} &=& \frac{(x^{\prime \, 0}_{l})^{2} e^{-2 g^{\mathrm{eff}}_{l}}}{1 + 
2 N^{\mathrm{eff}}_{l} + e^{-2 g^{\mathrm{eff}}_{l}}}\\
\tilde{A}_{p_{l}} &=& 1+\frac{e^{2 g^{\mathrm{eff}}_{l}}}{1+2 N^{\mathrm{eff}}_{l}}\\
\tilde{B}_{p_{l}} &=& \frac{x^{\prime \, 0}_{l} e^{2 g^{\mathrm{eff}}_{l}}}{1 + 
2 N^{\mathrm{eff}}_{l} + e^{2 g^{\mathrm{eff}}_{l}}} \\
\tilde{C}_{p_{l}} &=& \frac{(x^{\prime \, 0}_{l})^{2} e^{2 g^{\mathrm{eff}}_{l}}}{1 + 
2 N^{\mathrm{eff}}_{l} + e^{2 g^{\mathrm{eff}}_{l}}}
\end{eqnarray}
where the $\tilde{B}$ and the $\tilde{C}$ terms depend on the phase
$\phi$.  Finally, by exploiting the definition of the confluent
hypergeometric functions $U(a,b;z)$, the single-mode photon number
distribution can be written as:
\begin{equation}
\begin{aligned}
p(&n^{(l)} \vert \phi) = \frac{2 (-1)^{n}}{1+2 N_{l}^{\mathrm{eff}}}
e^{-2 (\tilde{C}_{x_{l}} + \tilde{C}_{p_{l}})}\sum_{k=0}^{n}
\sum_{j=0}^{k} \frac{2^{k}}{k!} \begin{pmatrix} n \\ k\end{pmatrix}
\begin{pmatrix} k \\ j \end{pmatrix} \\
&\times \frac{U[-j,1/2,-2 \tilde{A}_{x_{l}} (\tilde{B}_{x_{l}})^{2}]
  U[-k+j,1/2,-2 \tilde{A}_{p_{l}}
  (\tilde{B}_{p_{l}})^{2}]}{(\tilde{A}_{x_{l}})^{j+1/2}
  (\tilde{A}_{p_{l}})^{k-j+1/2}}
\end{aligned}
\end{equation}

\subsection{Derivative of the photon-number distribution and classical
  Fisher information}
In order to evaluate the classical Fisher information according to
Eqs. (\ref{eq:CFI_def_1}-\ref{eq:CFI_def_2}), we now need to evaluate
the derivative of the photon-number distribution $p(n^{(l)} \vert
\phi)$. The latter can be written in the following form
\begin{equation}
\begin{aligned}
p(&n^{(l)} \vert \phi) = \sum_{k=0}^{n} \sum_{j=0}^{k} \omega_{n,kj}
e^{-2 (\tilde{C}_{x_{l}} + \tilde{C}_{p_{l}})} \\
&\times \frac{U[-j,1/2,-2 \tilde{A}_{x_{l}} (\tilde{B}_{x_{l}})^{2}]
  U[-k+j,1/2,-2 \tilde{A}_{p_{l}}
  (\tilde{B}_{p_{l}})^{2}]}{(\tilde{A}_{x_{l}})^{j+1/2}
  (\tilde{A}_{p_{l}})^{k-j+1/2}}
\end{aligned}
\end{equation}
Here, $\omega_{n,kj}$ includes all the coefficients independent from
the phase $\phi$. The derivative of the photon-number distribution
$p(n^{(l)} \vert \phi)$ can then be written as the sum of three terms
\begin{equation}
\partial_{\phi} p(n^{(l)} \vert \phi) = \sum_{i=1}^{3} Dp_{i}(n^{(l)} \vert \phi)
\end{equation}
The term $Dp_{1}(n^{(l)} \vert \phi)$ presents the derivative of the 
exponential $e^{-2 (\tilde{C}_{x_{l}} + \tilde{C}_{p_{l}})}$, leading to:
\begin{equation}
Dp_{1}(n^{(l)} \vert \phi) = (-2) \partial_{\phi}(\tilde{C}_{x_{l}} + 
\tilde{C}_{p_{l}}) p(n^{(l)} \vert \phi)
\end{equation}
The terms $Dp_{2}(n^{(l)} \vert \phi)$ and $Dp_{3}(n^{(l)} \vert \phi)$ 
exploit the following relation involving the derivatives of the confluent 
hypergeometric functions:
\begin{equation}
\partial_{\phi} U[a,b,f(\phi)]= -a U[a+1,b+1,f(\phi)] \partial_{\phi}f(\phi)
\end{equation}
The remaining two terms can then be written as:
\begin{equation}
\begin{aligned}
D&p_{2}(n^{(l)} \vert \phi) = \frac{2 (-1)^{n}}{1+2
  N_{l}^{\mathrm{eff}}} e^{-2(\tilde{C}_{x_{H}}+\tilde{C}_{p_{H}})}
\sum_{k=0}^{n} \sum_{j=0}^{k} \begin{pmatrix} n \\ k \end{pmatrix}
\frac{2^{k}}{k!} \begin{pmatrix} k \\ j \end{pmatrix} \\ &\times
\frac{U[1-j,3/2,-2 \tilde{A}_{x_{l}} (\tilde{B}_{x_{l}})^{2}]
  U[-k+j,1/2,-2 \tilde{A}_{p_{l}}
  (\tilde{B}_{p_{l}})^{2}]}{(\tilde{A}_{x_{l}})^{j+1/2}
  (\tilde{A}_{p_{l}})^{k-j+1/2}} \\
& \times j (-4) \tilde{A}_{x_{l}} \tilde{B}_{x_{l}} (\partial_{\phi}
\tilde{B}_{x_{l}})
\end{aligned}
\end{equation}
and:
\begin{equation}
\begin{aligned}
D&p_{3}(n^{(l)} \vert \phi) = \frac{2 (-1)^{n}}{1+2
  N_{l}^{\mathrm{eff}}} e^{-2(\tilde{C}_{x_{H}}+\tilde{C}_{p_{H}})}
\sum_{k=0}^{n} \sum_{j=0}^{k} \begin{pmatrix} n \\ k \end{pmatrix}
\frac{2^{k}}{k!} \begin{pmatrix} k \\ j \end{pmatrix} \\ &\times
\frac{U[-j,1/2,-2 \tilde{A}_{x_{l}} (\tilde{B}_{x_{l}})^{2}]
  U[1-k+j,3/2,-2 \tilde{A}_{p_{l}}
  (\tilde{B}_{p_{l}})^{2}]}{(\tilde{A}_{x_{l}})^{j+1/2}
  (\tilde{A}_{p_{l}})^{k-j+1/2}} \\
& \times (k-j) (-4) \tilde{A}_{p_{l}} \tilde{B}_{p_{l}}
(\partial_{\phi} \tilde{B}_{p_{l}})
\end{aligned}
\end{equation}
Finally, the classical Fisher information can be evaluated according to:
\begin{equation}
I_{\phi} = \sum_{l=H,V} I_{\phi}^{(l)}
\end{equation}
where:
\begin{equation}
I_{\phi}^{(l)} = \sum_{n=0}^{\infty} \frac{(\sum_{i=1}^{3} Dp_{i}(n^{(l)} 
\vert \phi))^{2}}{p(n^{(l)} \vert \phi)}
\end{equation}

\section{Theory of the sensitivity of the protocol}\label{s:sensi}
\begin{figure}[ht!]
\centering
\includegraphics[width=0.35\textwidth]{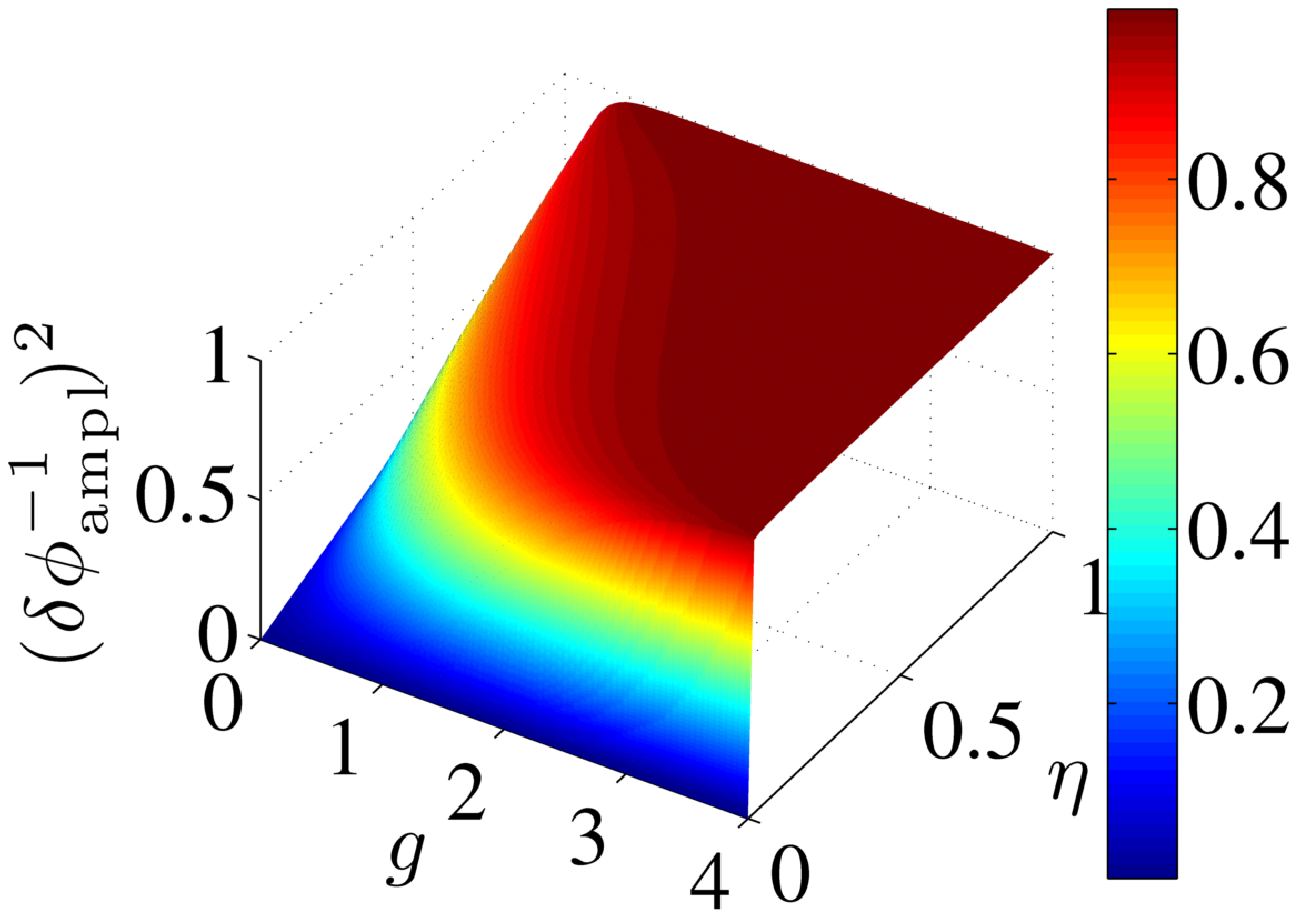}
\caption{Plot of $(\delta \phi_{\mathrm{ampl}}^{-1})^{2}$ as a function of the nonlinear gain $g$ of the amplifier and of the 
  detection efficiency $\eta$, with $\vert \beta \vert^{2} = 20$, normalized with respect to $I^{q}_{\mathrm{SQL}}$}
\label{fig:figS3}
\end{figure}

In this section we report the details of the calculation of the phase error 
$\delta \phi_{\mathrm{ampl}}$ for the proposed apparatus. It is convenient to work
in the Heisenberg picture. To this end, we need to consider the time
evolution of the field operators due to the OPA and of the loss map
$\mathcal{L}_\eta$. The latter is equivalent to the insertion of a
beam-splitter of transmissivity $\eta$ along the transmission path of
the field, seeded by the vacuum state in the other input port. By
combining the resulting equations for the time evolution of the
amplifier and of the beam-splitter we obtain the following expressions
for the field operators at the detection stage
\begin{eqnarray}
  \label{eq:in_out_plus}
  {c}^{\dag}_{H} &=& \sqrt{\eta} \left( {a}^{\dag}_{H} C + e^{- \imath
      \lambda} {a}_{H} S \right) - \imath \sqrt{1-\eta}
  {b}_{H}^{\dag}\\
\label{eq:in_out_minus}
{c}^{\dag}_{V} &=& \sqrt{\eta} \left( {a}^{\dag}_{V} C - e^{- \imath
    \lambda} {a}_{V} S \right) - \imath \sqrt{1-\eta} {b}_{V}^{\dag}
\end{eqnarray}
where ${b}^{\dag}_{H}$ and ${b}^{\dag}_{V}$ are the creation operators for the
second input port of the beam-splitter, $C = \cosh |g|$ and $S = \sinh
|g|$.  The chosen strategy to extract information on the phase shift
$\phi$ is to measure the output photon-number difference ${D} =
{c}^{\dag}_{H} {c}_{H} - {c}^{\dag}_{V} {c}_{V}$ and to extrapolate
the value of $\phi$ from the dependence of $\langle {D} \rangle$ on
it. By exploiting the expressions
(\ref{eq:in_out_plus}-\ref{eq:in_out_minus}) for the field operators,
the average of ${D}$ on the state ${\rho}_{\phi}^{\beta,g,\eta}$ is
\begin{equation}
  \langle {D} \rangle = \eta \vert \alpha \vert^{2} \xi \big[ \cos
  \phi (1+2 \overline{n}) + \cos(\phi+\lambda-2\theta) 2
  \sqrt{\overline{n}(1+\overline{n})} \big]
\end{equation}
To evaluate the resolution $\delta \phi$ on the estimated phase
according to standard estimation theory, we need to calculate the
fluctuations $\sigma(\langle {D} \rangle)$ on the detected signal.
Such quantity can be evaluated according to $\sigma^{2}(\langle {D}
\rangle) = \langle {D}^{2} \rangle - \langle {D} \rangle^{2}$. By
evaluating the average values $\langle ({c}^{\dag}_{H} {c}_{H})^{2}
\rangle$ and $\langle ({c}^{\dag}_{V} {c}_{V})^{2} \rangle$, we
obtain:
\begin{equation}
\sigma^{2}(\langle {D} \rangle) = \eta \big[ a(\overline{n},\eta) +
\cos \phi \cos(\phi+\lambda-2\theta) b(\overline{n},\eta) ]
\end{equation}
where:
\begin{eqnarray}
a(\overline{n},\eta) &=& 2 \overline{n} (1+\eta+2\eta
\overline{n})+\vert \alpha \vert^{2} \xi \big[ 1 + 2 \overline{n} +
\eta \overline{n} (6 + 8 \overline{n})\big]\nonumber\\\\
b(\overline{n},\eta) &=& 2 \sqrt{\overline{n}(1+\overline{n})} \vert
\alpha \vert^{2} \xi (1+\eta + 4 \eta \overline{n})
\end{eqnarray}
We note that both the signal and the fluctuations depend on the phase
difference between the coherent beam $\theta$ and the pump beam
$\lambda$. Finally, the resolution of this detection strategy can be
evaluated according to standard estimation theory as
\begin{eqnarray}
&&\delta \phi = {\sqrt{\sigma^{2}(\langle {D} \rangle)}}/{\left\vert
    \frac{\partial \langle D \rangle}{\partial \phi} \right\vert}=
\\
 &&\frac{\sqrt{a(\overline{n},\eta)+ \cos \phi
    \cos(\phi+\lambda-2\theta) b(\overline{n},\eta)}}{\vert \alpha
  \vert^{2} \sqrt{\eta} \xi \big\vert \cos \phi (1+2 \overline{n}) +
  \cos(\phi+\lambda-2\theta) 2 \sqrt{\overline{n}(1+\overline{n})}
  \big\vert}
\nonumber\\&&
\label{sensi}
\end{eqnarray}
Its optimal operating point is achieved for $\lambda - 2 \theta = 0$
and for a value of the actual phase of $\phi = \pi/2$, corresponding
to the steepest point of the signal $\langle {D} \rangle$. The error associated to the
phase estimation process in this optimal working point reads:
\begin{equation}
  \delta \phi_{\mathrm{ampl}} = \frac{a^{1/2}(\overline{n},\eta)}{\vert \alpha \vert^{2} \xi 
  \sqrt{\eta}(1 + 2 \overline{n} + 2 \sqrt{\overline{n}(1+\overline{n})})}.
\end{equation}
In Fig. \ref{fig:figS3} we report the value of $(\delta \phi_{\mathrm{ampl}}^{-1})^{2}$
normalized with respect to the SQL $I^{q}_{\mathrm{SQL}}$.
We note that for $\bar n\gg (2\eta)^{-1}$ and $|\beta|^2\gg1/2$
the QCR bound $\delta \phi\geqslant (M2|\beta|^2)^{-1/2}$ of the state
$|\Psi_\phi^\beta\>$ (before the amplification and the detector loss)
can be attained by our detection strategy.

The fact that $\delta \phi$ depends on the
parameter $\phi$ we want to estimate implies that the optimal regime 
$\delta \phi_{\mathrm{ampl}}$ can be achieved only by employing an
adaptive strategy, where some initial measurements are performed to
get an estimate of $\phi$ so that the apparatus can be employed in its
optimal working point around $\phi=\pi/2$. This is addressed in the next section.

\section{Adaptive protocol}\label{s:adaptive}
In this section we detail a simple two-stage adaptive scheme, where
first a rough estimate of the parameter $\phi$ is found, and then
this estimate is employed in a second high-resolution stage of the
protocol. 

\subsection{Bounds for a two-step adaptive protocol}

Let $\phi$ be the parameter we want to estimate (the phase) and assume
that it is encoded in two different families of states, i.e.  the
family $\{ \rho_{{{\phi}}} \}_{{\phi}}$ and the family $\{
\sigma_{{{\phi}}}\}_{{{\phi}}}$. For example, the first family can be
identified with the states of the system at the output of the
interferometer when no amplification is used. The second family
instead is identified as the the state at the output of the
interferometer when the amplifier is active and where we have set the
phase reference in such a way that the apparatus gives optimal
performances for ${{\phi}}=0$.  In what follows we will consider a two
stage estimation strategy in which {\em i)} first we perform $M_1$
measurements on the state $\rho_{{\phi}}$ of the first family to get a
preliminary estimation of ${{\phi}}$, and then {\em ii)} we perform
$M_2$ measurement on the state $\sigma_{{\phi}}$ of the second family
to improve our estimation (of course in the second stage we are
facilitated by the fact that we have already acquired some info on
${{\phi}}$).

Let then $\vec{x}= (x_1, x_2, \cdots)$ the data extracted from the
first set of measurement and ${{\phi}}_{ext}^{(M_1)}(\vec{x})$ the
estimation function we use to get the preliminary estimation of $
{{\phi}}$.  Using the quantum Cramer-Rao (QCR) bound we have
\begin{eqnarray}\label{eq1}
  \delta^2 {{\phi}}_1 = \sum_{\vec{x}} P_1(\vec{x}) [{{\phi}} -
  {{\phi}}_{ext}^{(M_1)}(\vec{x})]^2 \geqslant \frac{1}{M_1
    I^{q}_1({{\phi}})}\;,
\end{eqnarray}
where $P_1(\vec{x})$ are the probability of getting the outcomes
$\vec{x}$ when measuring $\rho_{{\phi}}^{\otimes M_1}$ and
$I^{q}_1({{\phi}})$ is the quantum Fisher info associated with the family
$\{ \rho_{{{\phi}}} \}_{{\phi}}$. For the sake of simplicity we assume
that $x_{ext}^{(M_1)}(\vec{x})$ is unbiased, i.e.
\begin{eqnarray}
  \sum_{\vec{x}} P_1(\vec{x})\big[{{\phi}} -
  {{\phi}}_{ext}^{(M_1)}(\vec{x})\big] =0\;,\label{eq2}
\end{eqnarray} 
(generalization to the general case are possible).

In the second stage of the estimation we use the family $\{
\sigma_{{{\phi}}}\}_{{{\phi}}}$ where we modify the way the phase is
mapped by rescaling it by ${{\phi}}_{ext}^{(M_1)}(\vec{x})$. This is
possible for instance by changing the initial phase reference which
effectively shifts the unknown phase ${{\phi}}$ to
$\chi={{\phi}}-{{\phi}}_{ext}^{(M_1)}(\vec{x})$: this is the new
parameter we wish to recover. In the second stage, we perform
measurements on $\sigma_{\chi}^{\otimes M_2}$ obtaining the data
$\vec{y}=(y_1, y_2, \cdots)$. We determine $\chi$ via the estimator
$\chi_{est}^{(M_2)}(\vec{y})$ which again we assume to be unbiased,
i.e.
\begin{eqnarray}
  \sum_{\vec{y}} P_2(\vec{y})\big[\chi- \chi_{ext}^{(M_2)}(\vec{y})\big]
  =0\;,
\end{eqnarray}
(here $P_2(\vec{y})$ is the probability of getting the outcomes
$\vec{y}$ when measuring $\sigma_{\chi}^{\otimes M_2}$).  The whole
process can be described hence by introducing a joint estimator
function
\begin{eqnarray}
\tilde{{{\phi}}}_{est}^{(M_1,M_2)}(\vec{x},\vec{y}) =
{{\phi}}_{ext}^{(M_1)}(\vec{x}) +\chi_{est}^{(M_2)}(\vec{y})\;.
\end{eqnarray} 
characterized by a probability distribution $P_1(\vec{x})
P_2(\vec{y})$ and which (by construction) is unbiased, i.e.
\begin{eqnarray}
\sum_{\vec{x},\vec{y}}P_1(\vec{x}) P_2(\vec{y}) \;
\tilde{{{\phi}}}_{est}^{(M_1,M_2)}(\vec{x},\vec{y}) = {{\phi}}\;.
\end{eqnarray} 
Let us now compute the variance of the error associated with such
estimator. Formally this is given by 
\begin{eqnarray}
  \delta^2 \tilde{{{\phi}}}  &=& \sum_{\vec{x},\vec{y}} P_1(\vec{x})
  P_2(\vec{y}) \;  [{{\phi}} -
  \tilde{{{\phi}}}_{ext}^{(M_1,M_2)}(\vec{x}, \vec{y})]^2 \nonumber\\ 
  &=&  \sum_{\vec{x}} P_1(\vec{x}) \left[ \sum_{\vec{y}}  P_2(\vec{y})
    \;  [{{\phi}} - \tilde{{{\phi}}}_{ext}^{(M_1,M_2)}(\vec{x},
    \vec{y})]^2 \right] \nonumber \\
  &=&  \sum_{\vec{x}} P_1(\vec{x}) \left[ \sum_{\vec{y}}  P_2(\vec{y})
    \;  [{{\phi}} -{{\phi}}_{ext}^{(M_1)}(\vec{x}) -
    \chi_{est}^{(M_2)}(\vec{y}) ]^2 \right] \nonumber \\
  &=&  \sum_{\vec{x}} P_1(\vec{x}) \left[ \sum_{\vec{y}}  P_2(\vec{y})
    \;  [\chi - \chi_{est}^{(M_2)}(\vec{y}) ]^2 \right] \nonumber
  \\\nonumber
  &\geqslant&  \sum_{\vec{x}} P_1(\vec{x}) \; \frac{1}{M_2 I^{q}_2(\chi)}
  \\&=&   \sum_{\vec{x}} P_1(\vec{x}) \; \frac{1}{M_2 I^{q}_2({{\phi}}
    -{{\phi}}_{ext}^{(M_1)}(\vec{x})
    )}\;,
\end{eqnarray}
where we used the QCR bound on the estimation of $\chi$ and where
$I^{q}_2(\chi)$ is the quantum Fisher info of the state $\sigma(\chi)$.
The above expression can now approximated by using the fact that for
sufficiently large $M_1$, $ {{\phi}}_{ext}^{(M_1)}(\vec{x})\simeq
{{\phi}}$, i.e. $\chi\simeq 0$. This allows us to expand $I^{q}_2(\chi)$
around $\chi=0$, i.e.
\begin{eqnarray}
&&I^{q}_2({{\phi}} -{{\phi}}_{ext}^{(M_1)}(\vec{x})) \simeq I^{q}_2(0) +
({{\phi}} -{{\phi}}_{ext}^{(M_1)}(\vec{x}) )\; I^{q \, \prime}_2(0) \nonumber\\
&& +({{\phi}}
-{{\phi}}_{ext}^{(M_1)}(\vec{x}))^2 \; I^{q \, \prime \prime}_2(0)/2 \;,
\end{eqnarray} 
which yields
\begin{eqnarray}
\delta^2 \tilde{{{\phi}}} &\simeq& \frac 1{M_2} \; \sum_{\vec{x}}
P_1(\vec{x}) \; {1}/\Big[I^{q}_2(0) + ({{\phi}}
  -{{\phi}}_{ext}^{(M_1)}(\vec{x}) ) \; I^{q \, \prime}_2(0) \nonumber\\&& +({{\phi}}
  -{{\phi}}_{ext}^{(M_1)}(\vec{x}))^2 \; I^{q \, \prime \prime}_2(0)/2\Big]
\nonumber \\
 &\simeq& \frac{1}{M_2 I^{q}_2(0)} \; \sum_{\vec{x}} P_1(\vec{x}) \; \Big[ 1 - 
 ({{\phi}} -{{\phi}}_{ext}^{(M_1)}(\vec{x}) ) \; \frac{I^{q \, \prime}_2(0)}{I^{q}_2(0)}  
 \nonumber \\
 && -  ({{\phi}} -{{\phi}}_{ext}^{(M_1)}(\vec{x}))^2 \; \frac{I^{q \, \prime \prime}_2(0)}{2 
 I^{q}_2(0)} 
\nonumber\\&&
+  ({{\phi}} -{{\phi}}_{ext}^{(M_1)}(\vec{x}) )^2 \; 
\left[\frac{I^{q \, \prime}_2(0)}{I^{q}_2(0)}\right]^2  \Big]
\nonumber \\
 &=& \frac{1}{M_2 I^{q}_2(0)} \;  \Big[ 1  - \delta^2 {{\phi}}_1  \; \Big( 
 \frac{I^{q \, \prime \prime}_2(0)}{2 I^{q}_2(0)} -\left[\frac{I^{q \, \prime}_2(0)}{I^{q}_2(0)}\right]^2\Big) 
 \Big]\;,\nonumber
\end{eqnarray}
where we used Eq.~(\ref{eq2}) and the definition of $ \delta^2
{{\phi}}_1$.  Suppose now that $I^{q}_2(\chi)$ achieves its maximum for
$\chi=0$ (this is what happens thanks to our new choice of reference).
This implies that $I^{q \prime}_2(0)=0$ and $I^{q \, \prime \prime}_2(0) \leqslant 0$. Therefore we
get
\begin{eqnarray}
&&\delta^2 \tilde{{{\phi}}} \geqslant  \frac{1}{M_2 I^{q}_2(0)} \;  \Big[ 1
+ \delta^2 {{\phi}}_1  \;  \frac{|I^{q \, \prime \prime}_2(0)|}{2 I^{q}_2(0)}
\Big]\nonumber\\&&
\geqslant 
\frac{1}{M_2 I^{1}_2(0)} \;  \Big[ 1  + \;  \frac{|I^{q \, \prime \prime}_2(0)|}{2 M_1 I^{q}_1({{\phi}}) 
I^{q}_2(0)} \Big] \;,
\end{eqnarray}
where in the last inequality we used the QCR bound~(\ref{eq1}).
Defining $M=M_1+M_2$ the total number of measurements, we can write
\begin{eqnarray} \label{ff1}
\delta^2 \tilde{{{\phi}}} \geqslant 
\frac{1}{(1-p) M I^{q}_2(0)} \;  \Big[ 1  + \;  \frac{|I^{q \, \prime \prime}_2(0)|}{2 p M
  I^{q}_{1}({{\phi}}) I^{q}_2(0)} \Big] \;,
\end{eqnarray}
with $p=M_1/M$ begin the fraction of measurement we employ in the
first step of the protocol. This equation provides the corrections to
the accuracy we get when we adopt the adaptive strategy.  \newline

{\bf Observation I:} It is worth comparing the above bound with the
accuracy one could get if instead of performing the preliminary step
one could have used all $M$ copies to perform only the estimation on
the states $\sigma_{{\phi}}$. In this case the resulting accuracy
would be $1/(M I^{q}_2({{\phi}}))$.  Do we gain something by going true
the adaptive result?  A positive answer would require
\begin{eqnarray} \label{fff}
\frac{1}{(1-p) M I^{q}_2(0)} \;  \Big[ 1  + \;  \frac{|I^{q \, \prime \prime}_2(0)|}{2p M
  I^{q}_1({{\phi}}) I^{q}_2(0)}    \Big] \leqslant \frac{1}{M
  I^{q}_2({{\phi}})}\;,
\end{eqnarray} 
which can be cast as
\begin{eqnarray} 
  \frac{p+A}{p(1-p)} \leqslant B\;,
\end{eqnarray}
with $B= I^{q}_2(0)/I^{q}_2({{\phi}})$ and $A= \frac{|I^{q \, \prime \prime}_2(0)|}{ 2 M
  I^{q}_1({{\phi}}) I^{q}_2(0)}$. Since by assumption $B\geqslant 1$ and
$A\geqslant 0$, one can easily verify that there are value of $p$
which allows one to obtain Eq.~(\ref{fff}) if $B$ is sufficiently
large.  \newline

{\bf Observation II:} For fixed $M$ we can optimize the
right-hand-side of Eq.~(\ref{ff1}) with respect to $p$. This yields
\begin{eqnarray} 
p_{opt} = \sqrt{A^2 +A} - A\;,
\end{eqnarray} 
(notice that this is and increasing function of $A$ which is always 
positive and smaller than 1/2 -- the latter being the asymptotic 
value reached for $A>>1$). Consequently we can write 
\begin{eqnarray} \label{fff2}
\delta^2 \tilde{{{\phi}}} &\geqslant &
\frac{1}{(1-p) M I^{q}_2(0)} \;  \Big[ 1  + \;  \frac{|I^{q \, \prime \prime}_2(0)|}{2p M
  I^{q}_1({{\phi}}) I^{q}_2(0)}    \Big]\nonumber\\&&
=\frac{1}{(1-p) M I^{q}_2(0)} \;  \Big[ 1
+ \;  \frac{A}{p}    \Big] \nonumber \\
& \geqslant &\frac{1}{M I^{q}_2(0)} \;  \frac{\sqrt{A^2 +A}}{(\sqrt{A^2
    +A} -A)(1+A-\sqrt{A^2 +A})}\;.\nonumber
\end{eqnarray} 
Now, for $M>>1$ we have that $A\rightarrow 0$. Therefore we can write 
\begin{eqnarray}  \delta^2 \tilde{{{\phi}}} &\geqslant&
  \frac{1}{M I^{q}_2(0)} \; [ 1+ 2 \sqrt{A}]\nonumber\\&=& \frac{1}{M
    I^{q}_2(0)} \; \Big[
  1+ \sqrt{\frac{2|I^{q \, \prime \prime}_2(0)|}{ M I^{q}_1({{\phi}}) I^{q}_2(0)}}\Big] \;.\label{qw}
\end{eqnarray} 
This implies that the resources $M_1$ employed in the first stage of
the protocol can be neglected, and the precision asymptotically
approaches the QCR of the second stage: the term with the square root
in \eqref{qw} is asymptotically negligible.

\subsection{Numerical simulation of a two-step adaptive protocol}

\begin{figure}[ht!]
\centering
\includegraphics[width=0.213\textwidth]{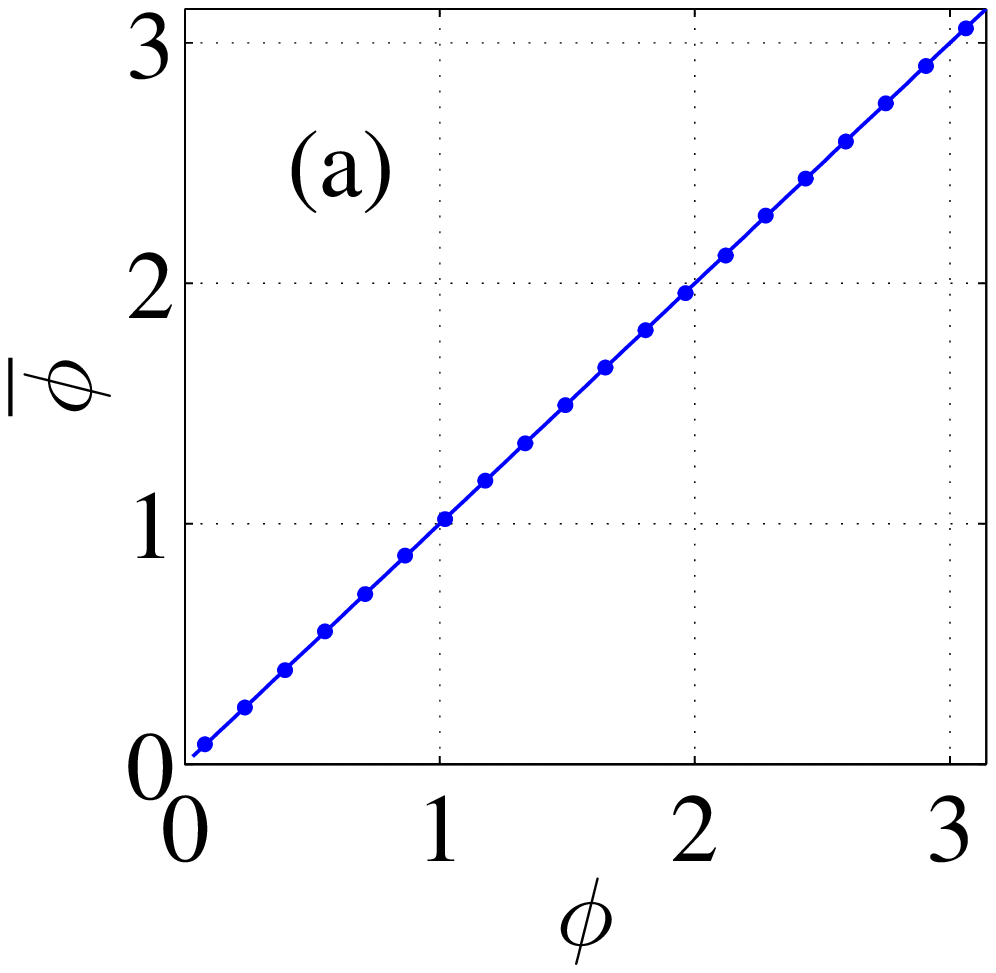}
\includegraphics[width=0.26\textwidth]{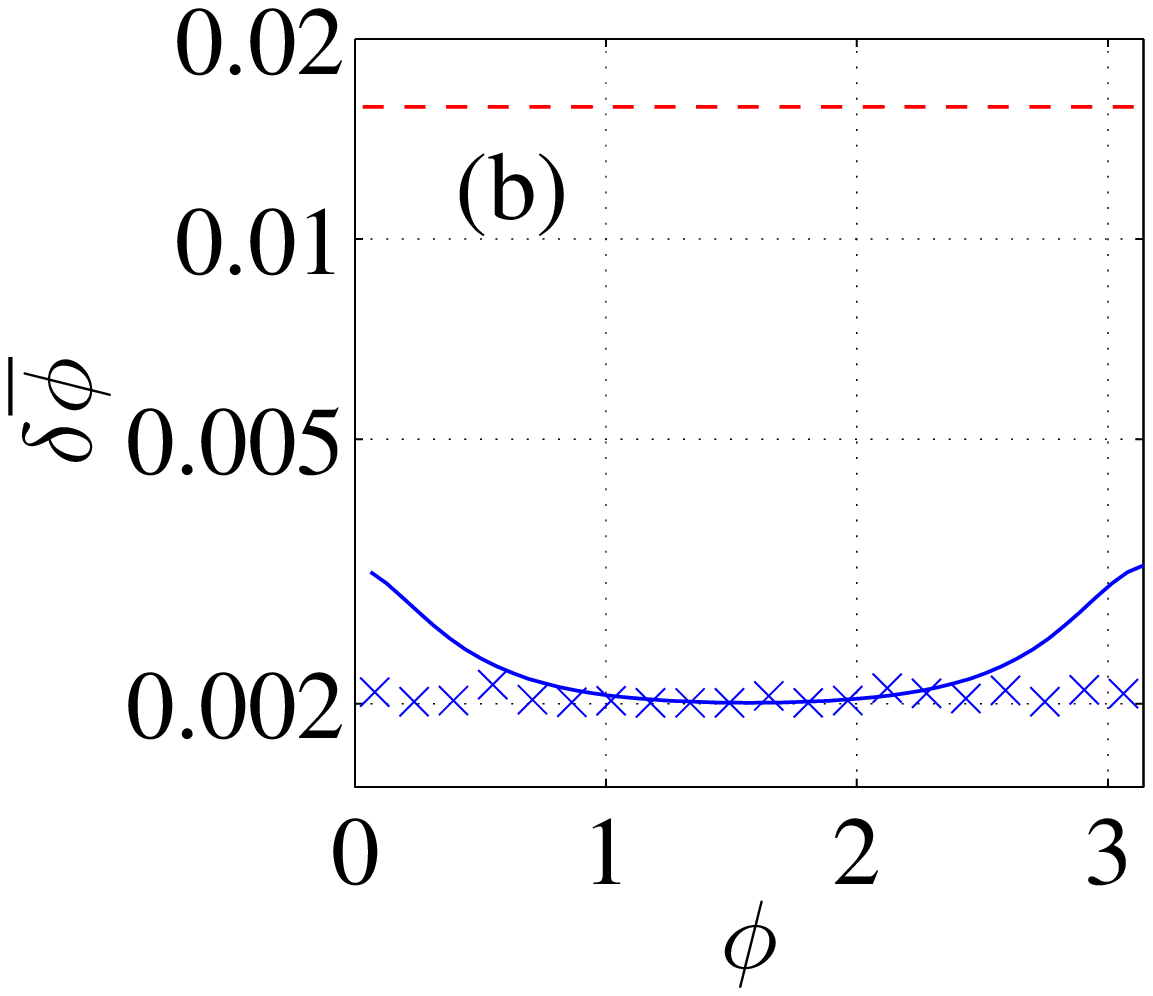}
\caption{Numerical simulation of a two step protocol for a phase estimation experiment 
  with the amplifier-based strategy, for $g=2$, $\vert \beta \vert^{2} = 4$, and $\eta = 10^{-2}$,
  with $M=10^{5}$ repeated measurements. (a) Estimated value $\bar{\phi}$ 
  and (b) corresponding error $\delta \bar{\phi}$ associated to the estimation process. 
  Points: numerical simulation. Blue solid line: classical Fisher information of the 
  amplifier-based protocol, which sets the bound for $\delta \phi$ without an adaptive strategy. 
  Red dashed line: classical Fisher information for a coherent 
  state protocol with the same parameters without the amplification strategy.}
\label{fig:figS4}
\end{figure}

Here we provide a numerical simulation of a two-step protocol tailored to reach the optimal performances,
given by the maximum of the classical Fisher information $I_{\mathrm{ampl}}$ in $\phi=\pi/2$ and 
$\lambda - 2 \theta=0$, for all the value of $\phi$. The two steps of the protocols are here described:
\begin{itemize}
\item[(I)] In a first step, a coherent probe state without the amplification-stage
(that is, by setting $g_{H}=g_{V}=0$) is adopted to obtain a rough estimate $\phi_{\mathrm{r}}$ of the phase.
\item[(II)] In a second step, the scheme is adjusted to the optimal working point by means of an additional 
phase shift $\psi$, which is tuned in order to set the overall phase of the interferometer to 
$\phi_{\mathrm{tot}} = \phi + \psi \simeq \pi/2$. Furthermore, the difference between the pump beam phase $\lambda$ and
the coherent state phase $\theta$ is set to $\lambda - 2 \theta=0$.
\end{itemize}
The data analysis on each step can be performed for instance by means of a Bayesian 
approach \cite{Pezz07}. In Fig. \ref{fig:figS4} we report the results of a numerical simulation
for $M=10^{5}$ repeated measurements. We observe that, for all values of the phase $\phi \in [0, \pi)$
the error $\delta \overline{\phi}$ reaches the maximum of the classical Fisher information, that is, $I_{\mathrm{ampl}}$
evaluted at $\phi = \pi/2$ and $\lambda - 2 \theta = 0$.

\section{Modeling the experiment}\label{s:exp}

Here we discuss the theoretical model for the analysis of the
experimental data of the protocol. In the implementation described in
the main paper, no phase stabilization is performed on the optical
path of the pump beam, hence the phase varies randomly at each
experimental run. To model such effect, an average on the phase
$\lambda$ with a uniform distribution $\mathcal{P}(\lambda) =
\frac{1}{2 \pi}$ must be performed on both the signal and the
fluctuations.  In this case, the average signal in the two
polarizations ${H}$ and ${V}$ is given by
\begin{eqnarray}
\overline{\langle {n}_{H} \rangle} &=& \eta \left[\overline{n} + 
\vert \alpha \vert^{2} \xi (1+2 \overline{n}) \cos^{2}(\phi/2) \right]\\
\overline{\langle {n}_{V} \rangle} &=& \eta \left[\overline{n} + 
\vert \alpha \vert^{2} \xi (1+2 \overline{n}) \sin^{2}(\phi/2) \right]
\end{eqnarray}
The average number of the count rates $\langle {D} \rangle$ is then
given by
\begin{equation}
\overline{\langle {D} \rangle} = \vert \alpha \vert^{2} \eta \xi \cos 
\phi (1+2 \overline{n})
\end{equation}
In the high losses regime investigated throughout the paper, the
number of photons effectively impinging on the detector is smaller
than one, since $\eta \langle {n}_{\pm} \rangle < 1$.  In this regime,
the single-photon counting process is described by a Poissonian
statistics. Hence, the fluctuation on the difference signal can be
evaluated as
\begin{equation}
  \sigma^{2}(\overline{\langle {D} \rangle}) =
  \sigma^{2}(\overline{\langle {n}_{H} \rangle}) +
  \sigma^{2}(\overline{\langle {n}_{V} \rangle}) = \overline{\langle
    {n}_{H} \rangle} + \overline{\langle {n}_{V} \rangle}
\end{equation}
By explicitly substituting the expressions for $\overline{\langle
  {n}_{H} \rangle}$ and $\overline{\langle {n}_{V} \rangle}$ we obtain
the following expression for the phase estimation error
\begin{equation}
\delta \phi = \frac{\sqrt{2 \overline{n} + \vert 
\alpha \vert^{2} \xi (1+2\overline{n})}}{\vert \alpha \vert^{2} \xi \sqrt{\eta} 
(1+2 \overline{n}) \vert \sin \phi \vert}
\end{equation}
The optimal point is achieved for $\phi = \pi/2$, where the error $\delta \phi$ is
\begin{equation}
\delta \phi_{\mathrm{exp}} = \frac{\sqrt{2 \overline{n} + \vert \alpha \vert^{2} \xi
    (1+2\overline{n})}}{\vert \alpha \vert^{2} \xi \sqrt{\eta} (1+2
  \overline{n})}
\end{equation}

\appendix

\section{Quantum Fisher Information}
\label{sec:Fisher_Information}
Here we briefly review the properties of the quantum Fisher
information for mixed states. Let us consider a family of states
${\sigma}_{\phi}$ depending on a parameter $\phi$. Such family of
states can be exploited to estimate the value of the parameter $\phi$.
In local estimation theory, the maximum amount of information that can
be extracted on the parameter $\phi$ with $M$ repeated measurements is
given by the QFI $I^{q}_{\phi}$. More specifically, the variance of any
estimator of the parameter $\phi$ satisfies the quantum Cramer-Rao
inequality:
\begin{equation}
\delta^{2} \phi \geq \frac{1}{M I^{q}_{\phi}}
\end{equation}
Here, $I^{q}_{\phi}$ represents the optimization of the classical Fisher 
information over all possible choice of the quantum measurement.
In general, the quantum Fisher information of the family of states 
${\sigma}_{\phi}$ is given by the following definition:
\begin{equation}
I^{q}_{\phi} = \mathrm{Tr} \big[ {\sigma}_{\phi} {L}_{\phi}^{2} \big]
\end{equation}
where ${L}_{\phi}$ is the symmetric logarithmic derivative of ${\sigma}_{\phi}$:
\begin{equation}
\partial_{\phi} {\sigma}_{\phi} = \frac{{L}_{\phi} {\sigma}_{\phi} + {\sigma}_{\phi} {L}_{\phi}}{2}
\end{equation}
By expressing the density matrix in terms of its spectral decomposition 
${\sigma}_{\phi} = \sum_{m} \sigma_{m} \vert \zeta_{m} \rangle \langle 
\zeta_{m} \vert$, the quantum Fisher information can be evaluated as 
\cite{Pari09}:
\begin{equation}
\label{eq:Fisher_def_mixed}
I^{q}_{\phi} = \sum_{p} \frac{(\partial_{\phi} \sigma_{p})^{2}}{\sigma_{p}} 
+ 2 \sum_{n,m} \epsilon_{n,m} \vert \langle \zeta _{m}\vert \partial_{\phi} 
\zeta_{n} \rangle \vert^{2}
\end{equation}
Here, $\partial_{\phi} \sigma_{p}$ is the derivative of the eigenvalues with 
respect to $\phi$, and $\vert \partial_{\phi} \zeta_{n} \rangle$ is the 
derivative of the eigenvectors written in a $\phi$-independent basis $\{ \vert k \rangle \}$:
\begin{equation}
\vert \partial_{\phi} \zeta_{m} \rangle = \sum_{k} (\partial_{\phi} \zeta_{mk}) \vert k \rangle
\end{equation}
Finally, the coefficient $\epsilon_{n,m}$ is given by the following expression:
\begin{equation}
\epsilon_{n,m} = \frac{(\sigma_{n}-\sigma_{m})^{2}}{\sigma_{n}+\sigma_{m}}
\end{equation}

\section{Mathematical relations}
\label{sec:Appendix_CV_relations}

In this Appendix we report some mathematical relations exploited in
the calculation of the Fisher information.

\textbf{\textsl{Thermal state.}} -- The thermal single-mode state is defined as:
\begin{equation}
{\rho}^{\mathrm{th}}(\overline{N}) = \frac{1}{1+\overline{N}} 
\sum_{n=0}^{\infty} \chi^{n} \vert n \rangle \langle n \vert 
\end{equation}
with $\chi = \overline{N}/(1+\overline{N})$, where $\overline{N}$ is the 
average number of photons of the state.

\textbf{\textsl{Lossy squeezed vacuum.}} -- The state generated by the 
action of a lossy channel on the squeezed vacuum state can be written 
as according to \cite{Aspa09}:
\begin{equation}
\label{eq:squeezed_vacuum_losses}
\mathcal{L}_{\eta} \left[ {S}({g}) \vert 0 \rangle \langle 0 
\vert {S}^{\dag}({g})\right] = {S}^{\dag}({g}^{\mathrm{eff}}) 
{\rho}^{\mathrm{th}}(N^{\mathrm{eff}}) {S}({g}^{\mathrm{eff}})
\end{equation}
The effective modulus of the squeezing parameter $g^{\mathrm{eff}}$ 
and the effective thermal noise $N^{\mathrm{eff}}$ take the form:
\begin{eqnarray}
\label{eq:g_tau_eff_1}
g^{\mathrm{eff}} &=& \frac{1}{4} \log(\frac{P}{M}) \\
N^{\mathrm{eff}} &=& \frac{-1+\sqrt{P M}}{2}
\end{eqnarray}
where:
\begin{eqnarray}
P &=& \eta e^{2 g} + 1 - \eta\\
\label{eq:g_tau_eff_4}
M &=& \eta e^{- 2 g} + 1 - \eta
\end{eqnarray}